\documentclass[twocolumn, preprintnumbers, superscriptaddress, nofootinbib]{revtex4-1}
\pdfoutput=1
\usepackage[pdftex]{graphicx}
\usepackage{amssymb,amsmath}
\usepackage[colorlinks,linktocpage]{hyperref}

\hypersetup{
    colorlinks=true,
    citecolor=blue
}

\begin{document}
\preprint{INR-TH-2020-013}
\title{Radio-emission of axion stars}

\author{D.G. Levkov}\email{levkov@ms2.inr.ac.ru}
\affiliation{Institute for Nuclear Research of the Russian Academy
  of Sciences, Moscow 117312, Russia}
\affiliation{Institute for Theoretical and Mathematical Physics, MSU,
  Moscow 119991, Russia} 
\author{A.G. Panin}\email{panin@ms2.inr.ac.ru}
\affiliation{Institute for Nuclear Research of the Russian Academy
  of Sciences, Moscow 117312, Russia}
\affiliation{Moscow Institute of Physics and Technology, 
  Dolgoprudny 141700, Russia}
\author{I.I. Tkachev}\email{tkachev@ms2.inr.ac.ru}
\affiliation{Institute for Nuclear Research of the Russian Academy
  of Sciences, Moscow 117312, Russia}
  \affiliation{Novosibirsk State University, Novosibirsk 630090,
    Russia}

\newcommand{\di}[1]{\textcolor{blue}{\bf #1}}  
\newcommand{\ig}[1]{\textcolor{red}{\bf #1}}  

\begin{abstract}
We study parametric instability of compact 
axion dark matter structures  decaying to radiophotons.
Corresponding objects~--- Bose (axion) stars, 
their clusters, and clouds of diffuse axions~---  form
abundantly in the postinflationary Peccei-Quinn scenario. We develop
general description of parametric resonance  incorporating
finite-volume effects, backreaction, axion velocities
and their (in)coherence. With additional coarse-graining, our
formalism reproduces kinetic equation for virialized axions interacting
with photons. We derive conditions for the parametric instability in each 
of the above objects, as well as in collapsing axion stars, evaluate
photon resonance modes and their growth  
exponents. As a by-product, we calculate stimulated emission of Bose
stars and diffuse axions, arguing that the former can give
larger contribution into the radiobackground. In the case of QCD
axions, the Bose stars glow and collapsing stars radioburst if the
axion-photon coupling exceeds the original KSVZ value by two orders of
magnitude. The latter constraint  is alleviated for several nearby
axion stars in resonance and absent for axion-like particles. Our
results show that the parametric effect may reveal itself in
observations, from FRB to excess  radiobackground.
\end{abstract}

\maketitle

\section{Introduction}
\label{sec:intro}
The QCD axion~\cite{Kim:2008hd} and similar
particles~\cite{pdg} are perfect dark
matter candidates~\cite{Sikivie:2006ni,Arias:2012az}: they are
motivated~\cite{Peccei:1977hh,Arvanitaki:2009fg} and have tiny
interactions~\cite{diCortona:2015ldu}, including coupling to the
electromagnetic field. But the same interactions --- alas~--- make the
axions ``invisible'' dictating overly precise detection
measurements~\cite{Irastorza:2018dyq,Armengaud:2019uso} and limiting
possible observational
effects~\cite{Arvanitaki:2014wva,Arza:2019nta,Foster:2020pgt}. 

Nevertheless, under certain conditions an avalanche of exponentially
growing photon number ${n_\gamma\propto   \exp\{2\mu_{\infty} t\}}$ can
appear in the axionic medium~\cite{Tkachev:1987cd}, with growth
exponent $\mu_{\infty}$ proportional to the axion-photon coupling and
axion field strength. This process is known as parametric resonance.
It occurs because the axions decay into photons which stimulate decays
of more axions. In the infinite volume parametric axion-photon
conversion is well understood, but does not occur during cosmological
evolution of the axion
field~\cite{Preskill:1982cy,Abbott:1982af,Alonso-Alvarez:2019ssa}. In compact
volume of size $L$ the avalanche appears if the photon
stimulates at least one axion decay as it passes the object
length~\cite{Tkachev:1987cd,Riotto:2000kh}. This gives
order-of-magnitude resonance condition, 
\begin{equation}
  \label{eq:D}
  \mu_{\infty} L \gtrsim 1\;.
\end{equation}
Unfortunately, apart from this intuitive estimate and brute-force numerical
computations~\cite{Kephart:1994uy,Tkachev:2014dpa,Hertzberg:2018zte,Arza:2018dcy,Sigl:2019pmj,Carenza:2019vzg,Chen:2020ufn,Wang:2020zur,Arza:2020eik},
no consistent quantitative theory of axion-photon conversion in finite-size
objects has been developed so far.

In this paper we fill this gap\footnote{This work is based on 
  presentations~\cite{Patras2018, Patras2019} at the Patras workshops
  where the main equations first appeared.} with a general, detailed, and
usable quasi-stationary approach to parametric resonance in a finite volume. Our
method works only for nonrelativistic axions,
but  accounts for their coherence, or its absence, axion velocities,
binding energy and gravitational redshift, backreaction of photons on
axions, and arbitrary volume shape. In the 
limit of diffuse axions it reproduces well-known axion-photon
kinetic equation, if additional coarse-graining is introduced.

Notably, the cosmology of  QCD axion~\cite{Preskill:1982cy,Abbott:1982af,Dine:1982ah}
provides rich dark matter structure at small
scales~\cite{Niemeyer:2019aqm}, with a host of potentially observable
astrophysical implications. Namely, in the postinflationary scenario
violent inhomogeneous evolution of the axion field during the QCD
epoch~\cite{Kolb:1993hw,Klaer:2017ond,Gorghetto:2018myk,Vaquero:2018tib,Buschmann:2019icd}
leads to formation of axion
miniclusters~\cite{Hogan:1988mp,Kolb:1993zz,Kolb:1994fi}~--- dense objects
of typical mass $10^{-13}\, M_{\odot}$ forming hierarchically bound
structures~\cite{Eggemeier:2019khm}. In the centers of miniclusters
even denser compact objects, the axion (Bose)
stars~\cite{Ruffini:1969qy,Tkachev:1986tr}, appear due to
gravitational kinetic
relaxation~\cite{Levkov:2018kau,Eggemeier:2019jsu}. Simulations
suggest~\cite{Levkov:2018kau,Vaquero:2018tib,Buschmann:2019icd,Niemeyer:2019aqm,Eggemeier:2019jsu}
that these objects are abundant in the Universe, though their
present-day mass is still under
study~\cite{Eggemeier:2019jsu}. Another example of dense
object formed by the QCD axions is a cloud around the superradiant
black hole~\cite{Arvanitaki:2010sy,Arvanitaki:2009fg,Stott:2018opm},
see also~\cite{Rosa:2017ury}.

Beyond the QCD axion, miniclusters~\cite{Arias:2012az} and Bose
stars~\cite{Schive:2014dra,Schive:2014hza,Amin:2019ums} can be formed by the
axion-like particles at very different length and mass scales. 

In this paper we derive precise conditions for parametric resonance in
the isolated axion stars, collapsing stars, their clusters, and in the
clouds of diffuse axions. We find unstable electromagnetic modes and their
growth exponents~$\mu$. Contrary to what naive infinite-volume
intuition might suggest, resonance in nonrelativistic compact
objects develops with $\mu \ll \mu_{\infty}$. As a result, in
many cases it glows in the stationary regime, burning an
infinitesimally small fraction of extra axions at every 
moment, to keep the resonance condition marginally broken.

Our calculations suggest three interesting scenaria with different
observational outcomes. In the first chain of events the axion-photon
coupling is high and the threshold for parametric resonance is
reached during growth of axion stars via Bose-Einstein
condensation. Then all condensing axions will be converted into
radio-emission with frequency equal to the axion half-mass. This paves
the way for indirect axion searches.

Second, at somewhat smaller axion-photon coupling, attractive
self-interactions of axions inside the growing stars may become important 
before the resonance threshold is reached. As a result,  the stars
collapse~\cite{Chavanis:2011zi,Zakharov12,Levkov:2016rkk}, shrink
and ignite the instability to photons on the way. Alternatively,
several smaller axion stars may come close, suddenly meeting the
resonance 
condition~\cite{Hertzberg:2018zte}. In these cases a short and
powerful burst of radio-emission appears. 

Amusingly, powerful and unexplained Fast Radio Bursts (FRB) are
presently observed in the sky~\cite{Petroff:2019tty}. It is tempting
to relate them to parametric resonance in collapsing axion
stars~\cite{Tkachev:2014dpa} and see if the main characteristics
can be met. 

In the third, most conservative scenario all Bose stars are far away
from the parametric resonance. Nevertheless, the effect of stimulated
emission turns them into powerful radioamplifyers of ambient 
radiowaves at the axion half-mass frequency. We compute 
amplification coefficients for the Bose stars  and diffuse axions and
find that realistically, stimulated emission of the stars may give larger
contribution into the radiobackground.

In Sec.~\ref{sec:BS} we introduce nonrelativistic approximation for
axions and review essential properties of Bose stars. General
description of parametric resonance in finite volume is
developed in  Sec.~\ref{sec:general-formalism}. In
Sec.~\ref{sec:reson-stat-matt} it is applied to radio-emission of
static axion stars, their pairs, and amplification of ambient
radiation. In Secs.~\ref{sec:diffuse-axions}
and~\ref{sec:moving-axions} we study resonance in diffuse axions and
consider the effect of moving axions / axion stars, in particular,
resonance in collapsing stars. Concluding remarks are given in
Sec.~\ref{sec:discussion}.


\section{Axion stars}
\label{sec:BS}

The diversity of compact objects in axion cosmology offers many
astrophysical settings
where the
parametric resonance may be expected. One can consider static Bose
stars, collapsing, moving, or tidally disrupted stars, even axion
miniclusters. To describe all this spectrum in one go, we implement 
two  important approximations. 

First we describe axions by  the classical field $a(t,\, {\bf x})$ satisfying
\begin{equation}
  \label{eq:34}
  \Box a + {\cal V}'(a) = 0\;.
\end{equation}
This is valid at large occupation numbers.
Interaction with the gravitational field in Eq.~(\ref{eq:34}) is hidden  in the
covariant derivatives, and the scalar potential
\begin{equation}
{\cal V} = \frac{m^2}{2}\, a^2  - \frac{g_4^2 m^2}{4! \, f_a^2}\, a^4  + \dots
\label{eq:aPot}
\end{equation}
includes mass $m$ and quartic coupling $(g_4 m/f_a)^2$.
Self-interaction of the QCD axion is attractive: ${f_a  \simeq
    (75.5~\rm MeV)^2}/m$ and $g_4  \simeq
  0.59$~\cite{diCortona:2015ldu}.  Axion-like particles may have
  ${g_4 \simeq 0}$.

Second, we work in nonrelativistic approximation,
\begin{equation}
  \label{ansatz}
  a = \frac{f_a}{\sqrt{2}}\, \left[\psi(t,\, {\bf x})\,\mathrm{e}^{-i
      m t} + \mathrm{h.c.}\right]\;,  
\end{equation}
where $\psi$ slowly depends on space and time. Namely, if $\lambda$ is
the typical wavelength of axions,
\begin{equation}
  \label{nonrel}
  \partial_t \psi \sim \psi/ m\lambda^2 \;,  \qquad
  \partial_{\bf x} \psi \sim \psi/\lambda\;, \qquad \lambda m \gg 1\;.
\end{equation}
 In this approximation Eq.~(\ref{eq:34}) reduces to nonlinear
Schr\"odinger equation,
\begin{equation}
  \label{eq:36}
  i \partial_t \psi = -\frac{\Delta}{2m}\, \psi + 
  m \left(\Phi - \frac{g_4^2}{8} |\psi|^2 \right)\psi\;,
\end{equation}
where $\Phi$ is a nonrelativistic gravitational potential solving
the Poisson equation
\begin{equation}
  \label{eq:40}
  \Delta \Phi = 4 \pi \rho/M_{pl}^2\;, 
\end{equation}
and $\rho = m^2 f_a^2 |\psi|^2$ is the mass density of axions.

Note that the method of this paper is applicable only if both of
the above conditions are satisfied: the axions are nonrelativistic   and
they have large occupation numbers. Dark matter axions meet these
requirements, except under extreme conditions.

A central object of our study is a Bose (axion) star, a  stationary
solution to the  Schr\"odinger-Poisson system
\begin{equation}
\label{eq:BS}
 \psi =\mathrm{e}^{-i\omega_s t} \psi_{s}(r)\;, \qquad\qquad \Phi = \Phi_{s}(r), 
\end{equation}
where $\omega_s<0$ is the binding energy of axions and $r$ is the
radial coordinate. Physically, Eq.~(\ref{eq:BS}) describes
Bose-Einstein condensate of axions occupying a ground state in the
collective potential well $\Phi_s(r)$. This object is coherent: the
complex phase of $\psi_s$ does not depend on space and time. Below  we
consider parametric resonance in stationary and colliding stars. 

Notably, the axion stars with the critical mass
\begin{equation}
\label{eq:BSC}
M_{cr} \simeq 10.2\; \frac{f_a M_{pl}}{m g_4 }
\end{equation}
and heavier stars are unstable~\cite{Chavanis:2011zi}. In
this case attractive self-interaction in Eq.~(\ref{eq:36}) overcomes
the quantum pressure and 
the star starts to shrink developing huge axion
densities in the center~\cite{Levkov:2016rkk}. 
We will see that this may trigger explosive parametric instability. 


\section{General formalism}
\label{sec:general-formalism}

\subsection{Linear theory}
\label{sec:3}
In this Section we construct general quasi-stationary theory for narrow parametric 
resonance of radiophotons in the finite volume filled with axions. This
technique was first developed and presented
in~\cite{Patras2018,Patras2019}. In contrast to the  resonance in the infinite
volume~\cite[]{Preskill:1982cy,Abbott:1982af,Tkachev:1986tr} which
universally leads to the Mathieu  equation, the finite-volume one is
described by the eigenvalue problem with a rich variety of solutions. 

Consider  Maxwell's equations\footnote{We disregard
  gravitational interaction of photons. It will be restored below.}
for the electromagnetic potential $A_\mu$
in the axion background~$a(t, {\bf x})$, 
\begin{equation}
  \label{equation}
  \partial_\mu \left (F_{\mu\nu} +
  g_{a\gamma\gamma} a \tilde{F}_{\mu \nu} \right ) = 0\;,
\end{equation}
where $F_{\mu\nu} = \partial_\mu A_{\nu} - \partial_\nu A_\mu$,
${{\tilde F}_{\mu \nu} \equiv 
  \epsilon_{\mu\nu\lambda\rho}F_{\lambda\rho}/2}$, and $g_{a\gamma 
  \gamma}$ is the standard axion-photon coupling. Below we also use
dimensionless coupling $g' \equiv f_a g_{a\gamma \gamma}/2^{3/2}$.

In the infinite volume one describes the resonance in the plain-wave
basis for the electromagnetic
field~\cite{Preskill:1982cy,Abbott:1982af,Tkachev:1986tr}, while
the axion star suggests spherical
decomposition~\cite{Hertzberg:2018zte}. We want to develop general 
formalism, and simpler at the same time, usable in a large variety of 
astrophysical settings.

We therefore introduce two simplifications. First, the photons travel straight, with
light-bending effects being subdominant in the axion background,
cf.~\cite{Blas:2019qqp,McDonald:2019wou}. Thus,  
the parametric resonance develops almost independently along different 
directions. Second, we consider non-relativistic axions decaying
into photons of frequency $\omega_{\gamma}\approx m/2$ with a very
narrow spread.

This suggests decomposition in the
gauge $A_0=0$,  
\begin{equation}
  \label{eq:1}
  A_i = \int d\boldsymbol{n}\; C_i^{(\boldsymbol{n})}(t,{\bf x})\,
  \mathrm{e}^{im(\boldsymbol{n}{\bf x}+t)/2} + \mathrm{h.c.}\;,
\end{equation}
where $i = \{x,\, y,\, z\}$  and the integral runs over all unit
vectors $\mathbf{n}$. The amplitudes $C_{i}^{(\boldsymbol{n})}$
include photon frequency spread. Hence, they weakly
depend on space and time, 
\begin{equation}
  \label{eikonal}
  \partial_{t,{\bf x}} |C_i^{(\boldsymbol{n})}| \sim
  \lambda^{-1}|C_i^{(\boldsymbol{n})}| \;,
  \qquad \lambda m \gg 1 \;,
\end{equation}
where $\lambda^{-1}$ is the typical momentum of  axions
in Eq.~\eqref{nonrel}.

Using Eq.~\eqref{eq:1}, one finds that in the eikonal
limit~\eqref{eikonal} the field equation~\eqref{equation} couples only
the waves moving in the opposite, i.e.\ $+ \boldsymbol{n}$ and 
$-\boldsymbol{n}$, directions. As a result, identical and independent
equations are produced for every pair of directions.  This is
manifestation of the simple fact that the axions decay into two
back-to-back photons. 

Indeed, leaving one arbitrary direction $z = (\boldsymbol{n}{\bf  x})$
and its counterpart $-z$, we obtain the ansatz that passes the
field equation~\cite{Patras2018,Patras2019}, 
\begin{equation}
  \label{plane}
  A_i = C_i^+ \,  \mathrm{e}^{im(z+t)/2} +
  C_i^- \, \mathrm{e}^{im(z-t)/2} + \mathrm{h.c.}\;,
\end{equation}
where the shorthand notations ${C_{i}^+(t,\, {\bf x}) =
  C_i^{(\boldsymbol{n})}}$ and $C_{i}^-(t,\, {\bf x}) =
[C_i^{(-\boldsymbol{n})}]^*$ are introduced. Namely, substituting
Eq.~(\ref{plane}) into Eq.~\eqref{equation} and using
approximations~\eqref{eikonal},~\eqref{nonrel}, 
we arrive to the closed system, 
\begin{subequations}
  \label{syst}
\begin{align}
  \label{syst1}
  &\partial_t C^+_x = \partial_z C^+_x + {i g'} \, m \psi^* C^-_y  \;,\\
  \label{syst2}
  &\partial_t C^-_y = - \partial_z C^-_y  - {i g'}\,  m \psi C^+_x \;.
\end{align}
\end{subequations}
 The other two physical polarizations satisfy the same
 equations with $C^+_x \to C^+_y$ and $C^-_y \to   -C^{-}_x$, while
 the longitudinal part is fixed by the Gauss law ${C_{z}^{\pm} = 2i
     \partial_\alpha C_\alpha^\pm/m}$; here and below $\alpha =
 \{x,\, y\}$. Overall, we have four  amplitudes $C_{\alpha}^\pm$
 representing two photon polarizations 
 propagating in the $+z$  and $-z$ directions.

  Equations~(\ref{syst}) should be solved for every orientation
    of $z$ axis, in search for the growing instability modes. After
    that the modes can be superimposed  in  Eq.~(\ref{eq:1}) or,
    practically,  only the one with the largest exponent can be kept.

In spherically symmetric Bose star all directions are equivalent and
description simplifies --- we have to study only one
direction. Notably, in this case one can derive Eqs.~\eqref{syst}
using spherical decomposition, see
Appendix~\ref{sec:spherical-symmetry}. 

There is a residual hierarchy in Eqs.~\eqref{syst} related to small
axion velocities $v \sim (m\lambda)^{-1}\ll 1$. Indeed, the   
nonrelativistic background evolves slowly, 
${\partial_t \psi \sim \psi/m\lambda^2}$, while the electromagnetic
field changes fast, ${\partial_{t} C  \sim C/\lambda}$. Thus,
equations for $C$ can be solved with adiabatic
ansatz,
\begin{equation}
  \label{eq:4}
  C_{i}^{\pm} = \mathrm{e}^{\int\limits^t \mu(t') \, dt'} \, c_{i}^{\pm}(t,
  \,{\bf x})\;,
\end{equation}
where the complex exponent $\mu$ and quasi-stationary amplitudes
$c_{i}^{\pm}$ evolve on the same timescales $m\lambda^2$ as 
$\psi$. Corrections to the adiabatic evolution~(\ref{eq:4}) become
exponentially small as $v \to 0$.

Using representation~(\ref{eq:4}) in
Eqs.~(\ref{syst}) and ignoring time derivatives of
$\mu$ and $c_{i}^{\pm}$, we finally obtain the eigenvalue
problem~\cite{Patras2018,Patras2019},
\begin{subequations}
  \label{sys}
\begin{align}
  \label{sys1}
  &\mu c^+_x = \partial_z c^+_x   +  {i g' }\, m \psi^*\, c^-_y\;, \\
  \label{sys2}
  &\mu c^-_y = - \partial_z c^-_y   - {ig'} \, m \psi \,c^+_x\;,
\end{align}
\end{subequations}
where equations for the two remaining amplitudes are again obtained by
$c^+_x \to c^+_y$ and ${c^-_y \to -c^{-}_x}$.

If the axions live in a finite region  and no electromagnetic waves come
from infinity, one imposes boundary conditions
\begin{equation}
  \label{bound}
  \left. c^+_\alpha \right |_{z \to +\infty} = \left. c^-_\alpha \right|_{z \to -\infty} 
  = 0\; ,
\end{equation}
see Eq.~(\ref{plane}).

The spectral problem~\eqref{sys} determines the electromagnetic modes
$\{c^+_x,\, c^-_y\}$ and their growth exponents $\mu$. The latter are
not purely imaginary at $\psi\ne 0$  because $2\times 2$ operator in
the right-hand side of Eqs.~(\ref{sys}),  is not anti-Hermitean. That
is why in certain cases  resonance instabilities~--- modes with
$\mathrm{Re}\, \mu>0$ satisfying the boundary conditions (\ref{bound})~---
appear. 

It is worth discussing two parametrically small corrections to
Eqs.~(\ref{syst}), \eqref{sys}. First, derivatives with respect to $x$
and $y$ are absent in these systems: they appear only in the next,
$(m\lambda)^{-1}$ order, determining the section of the resonance
  ray in the $(x,\, y)$ plane. If needed, they can be recovered with the substitution 
\begin{equation}
  \label{eq:35}
  \partial_z  \to  \partial_z - \frac{i}{m} (\partial_x^2 + \partial_y^2)\;,
\end{equation}
to solve the spectral problem in three dimensions.

If  the axion distribution is not
spherically-symmetric, one expects that the resonance
ray is narrow in the $(x,\, y)$ plane.\footnote{Similarly, one
    can compute narrow resonance rays around every direction in
    spherical axion star, and then combine them in~\eqref{eq:1}.}  Indeed, according to the
leading-order equations~\eqref{syst} electromagnetic field grows with
different exponents $\mu$ at different $x$ and $y$. This means that
wide wave packets shrink around the resonance line until the quantum
pressure  \eqref{eq:35} becomes relevant.

Second, direct interaction of photons with nonrelativistic
  gravitational field can be included in Eqs.~(\ref{syst}) by
  changing\footnote{Here $m\Phi$ accounts for the gravitational
    evolution of photon four-momentum.}
  \begin{equation}
    \label{eq:2}
    \partial_z C_{\alpha}^{\pm} \to \left(\partial_z + im \,
    \Phi\right) C_{\alpha}^{\pm}\;.
  \end{equation}
However, one immediately rotates this contribution away, 
  ${C_{\alpha}^\pm \to \exp\{-im\int^z dz' \, \Phi(z')\} \, C_{\alpha}^{\pm}}$,
with remaining corrections suppressed by~$(m\lambda)^{-2}$.

As an illustration, consider static homogeneous axion field $\psi$
in the infinite volume. Quasi-stationary equations (\ref{sys}), 
in this case give $\partial_z c_{\alpha}^\pm = 0$ and
time-independent $\mu = \mu_{\infty}$ of the form
\begin{equation}
\mu_{\infty} = g'm\, |\psi| >0\;.
\label{muRsimpe}
\end{equation}
Thus, electromagnetic amplitudes in Eq.~\eqref{eq:4} grow
exponentially with time indicating parametric resonance. Expression
\eqref{muRsimpe} reproduces well-known infinite-volume growth 
rate~\cite{Preskill:1982cy,Abbott:1982af,Tkachev:1986tr,Tkachev:1987cd,Kephart:1994uy,Riotto:2000kh,Hertzberg:2018zte,Sigl:2019pmj} of the axion-photon resonance.

\subsection{Nonlinear stage}
\label{sec:nonlinear-stage}

Backreaction of photons on axions can be easily 
incorporated in the Schr\"odinger-Poisson system (\ref{nonrel}). To
this end one substitutes the nonrelativistic ansatz (\ref{ansatz}),
(\ref{plane}) into the equation for the axion field, 
\begin{equation}
  \label{eq:5}
  \Box a +  {\cal V}'(a)  = - \frac{g_{a\gamma\gamma}}{4} \,F_{\mu\nu}\tilde{F}_{\mu\nu}\;,
\end{equation}
and omits higher derivatives of
$\psi$ and $C$. This gives,
\begin{multline}
  \label{eq:8}
  i \partial_t \psi = -\frac{\Delta \psi}{2m}  + m \Phi\psi -
  \frac{mg_4^2}{8} \,|\psi|^2\psi\\
  - \frac{m g'} {f_a^2}\, \epsilon_{\alpha \beta}\,
  C_{\alpha}^- C_{\beta}^{+*},
\end{multline}
where the backreaction is represented by the new term\footnote{ If several directions 
    in the transform~\eqref{eq:1} are essential,  a combination of backreaction terms appears here. If
    spherical decomposition is used for the isolated axion star, these
    terms come with factors $r^{-2}$ in front, see
    Appendix~\ref{sec:spherical-symmetry}.} in the
right-hand side.

Let us show that the last term in the above equation changes the mass
$M = m^2 f_a^2 \int d^3 \boldsymbol{x}\, |\psi|^2$ of the axion
cloud. Indeed, taking the time derivative of $M$ and using
Eq.~(\ref{eq:8}),  we obtain energy conservation
law, 
\begin{equation}
  \label{eq:14}
  \partial_t M = J_{in}   - \int dx dy \, F_{a\to \gamma\gamma}\;,
\end{equation}
where $J_{in} = - mf_a^2 \int d^2\sigma^i \, \mathrm{Im} (\psi^*
\partial_i \psi)$ is the mass of axions entering the system per unit
time and
\begin{equation}
  \label{eq:9}
  F_{a\to \gamma\gamma} = 2 {m^3 g'}  \int dz   \,
  \epsilon_{\alpha \beta} \, \mathrm{Im} (\psi^* C_\alpha^- C_\beta^{+*})
\end{equation}
is the flux of produced photons. Below we will also use
the electromagnetic Poynting fluxes at infinity,
\begin{equation}
  \label{eq:22}
  F_{\gamma}^\pm = \mp m^2(|C_x^\pm|^2 + |C_y^{\pm}|^2)/2\;.
\end{equation}
In conjunction with Eqs.~\eqref{syst} this gives conservation law for the
electromagnetic energy, ${\partial_t E_{\gamma} = \int dx dy \,
  \left(F_{a\to \gamma\gamma} - [F_{\gamma}^+ +
    F_{\gamma}^{-}]_{z=-\infty}^{z=+\infty}\right)}$.

To conclude, one can numerically solve Eq.~(\ref{eq:8}) together
with Eqs.~(\ref{syst}) and watch the axions burn abundantly. 

\section{Static coherent axions}
\label{sec:reson-stat-matt}

\subsection{Condition for resonance}
\label{sec:glowing-axion-stars}

For a start, consider the case when the axions in a finite volume
are coherent and  do not move. A notable example of this situation is
a static axion star.

\begin{figure}
  \includegraphics{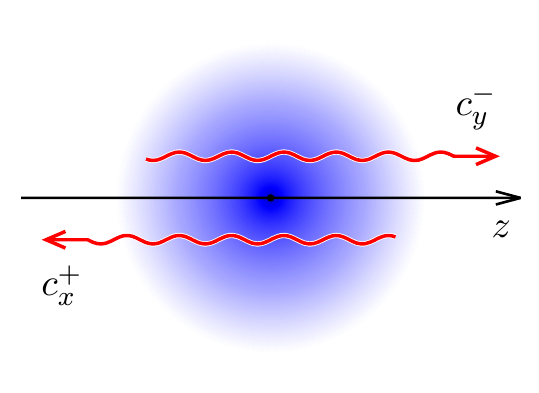}
  \caption{Parametric resonance in the axion star.\label{fig:bose_star}}
  \label{fig:resonance_illustrate}
\end{figure}

Parametric resonance in this setup is presently understood at the
qualitative level~\cite{Tkachev:1987cd,Tkachev:2014dpa,Hertzberg:2018zte}. Indeed,  
photons passing through the axions stimulate their decays ${a\to
  2\gamma}$. The photon flux  grows exponentially,
${F_{\gamma}\propto \exp\{2\mu_{\infty} t\}}$, and the secondary flux of backward-moving
photons appears. After the original photons escape the region with axions, 
stimulated decays continue in the secondary flux
moving in the opposite direction, etc, see
Fig.~\ref{fig:resonance_illustrate}. Overall, the back-and-forth
motion inside the axion cloud accumulates photons at every pass if
Eq.~\eqref{eq:D} is valid, i.e.
\begin{equation}
  \label{muRsimple}
\mu_{\infty} L = g'm |\psi| L \, \gtrsim 1\;,   
\end{equation}
where $L$ is the typical size of the cloud.

Our equations~(\ref{syst}) reflect the same
physics. Namely, consider the localized wave packet $C^{-}_y(t,\, {\bf 
  x})$ going through axions in the $+z$ direction. Due to
Eq.~(\ref{syst1})  it creates the packet $C^+_x$ with the opposite
group velocity which, in turn, produces $C^{-}_y$, etc. The photon
flux grows exponentially during this process if unstable modes with
$\mathrm{Re}\, \mu \geq 0$ are present.

Axion velocities are related to the complex phase of the  field,
\begin{equation}
  \label{eq:6}
  v_i = m^{-1}\partial_i\, \mathrm{arg}\, \psi\;.
\end{equation}
In this section we assume that $\psi$ is real up to a constant
phase which can be absorbed into redefinition of $c_{x}^{\pm}$ in
Eqs.~(\ref{sys}). This means that the axions are static and coherent.

In particular, the phase factor $\exp\{-i\omega_s t\}$ of the
Bose star field (\ref{eq:BS}) disappears from the electromagnetic
equations after replacing $C_i^{\pm} \to C_i^{\pm} \mathrm{e}^{\pm
  i\omega_s t/2}$. Then the total binding energy $\omega_s$ of axions
  inside the star does not
destroy the resonance, but slightly shifts its central frequency to 
\begin{equation}
\label{eq:omgamma}
\omega_\gamma = (m + \omega_s)/2\;,
\end{equation}
see Eq.~(\ref{plane}). Note that misconceptions regarding resonance
blocking by gravitational and self-interaction energies still
exist in the literature, e.g.~\cite{Wang:2020zur}.

At real $\psi$ the semiclassical eigenvalue problem~\eqref{sys} has
two types of solutions. First, delocalized modes penetrate into the
asymptotic regions $z\to \pm \infty$, where $\psi=0$ and 
$c_i^{\pm} \propto \exp(\pm \mu z)$. The  exponents $\mu$ of these modes
are purely imaginary, or their profiles would be
unbounded. Physically, the delocalized modes represent electromagnetic waves
coming from infinity. Second, there may exist
localized modes satisfying the boundary  
conditions~(\ref{bound}). They behave well at infinity if
$\mathrm{Re}\, \mu \geq 0$. In addition, we prove in 
Appendix~\ref{sec:prop-sympl-oper} that at real $\psi$ the exponents
$\mu$ of these modes are real. The localized modes represent
resonance instabilities.

In practical problems the resonance is not present in matter from the
very beginning but appears in the course of nonrelativistic
evolution. For example, the Bose stars form in slow 
galactic~\cite{Schive:2014dra,Schive:2014hza,Veltmaat:2016rxo,Veltmaat:2018dfz}
or minicluster~\cite{Kolb:1993zz,Eggemeier:2019jsu} collapses, or
afterwards in kinetic relaxation~\cite{Levkov:2018kau}, then 
grow kinetically at turtle-slow rates~\cite{Levkov:2018kau,Eggemeier:2019jsu,Veltmaat:2019hou}. Their subsequent evolution is also essentially nonrelativistic~\cite{Schwabe:2016rze,Schive:2019rrw}.

At some point of quasi-stationary evolution one of purely imaginary
eigenvalues $\mu$  may become real, and the parametric resonance
develops. Let us discuss the borderline situation when the very first
localized mode has $\mu=0$. The solution in  this case
is~\cite{Patras2018,Patras2019},
\begin{equation}
  \label{sol}
  c_x^+ = A \cos D(z)\;,\quad c_y^- = -i A \sin D(z)\;,
\end{equation}
where $A$ is a constant amplitude and
\begin{equation}
  \label{S}
  D(z) = {g'm }  \int\limits_{-\infty}^z dz' \, \psi(z')\;.
\end{equation}
Integration in Eq.~(\ref{S}) runs along the
arbitrary-oriented $z$-axis.

The solution~\eqref{sol} satisfies the boundary conditions (\ref{bound})
if $D_{\infty} \equiv D(+\infty) = \pi/2$. At larger values of this integral the
instability mode with positive $\mu$ exists. Thus, a precise condition
for the parametric resonance along a given $z$-axis is 
\begin{equation}
  \label{Sinf}
  D_{\infty} \equiv {g'm} \int\limits_{-\infty}^{+\infty} \psi(z) \,dz \geq \frac{\pi}{2}\;.
\end{equation}
This concretizes the order-of-magnitude
estimate~(\ref{muRsimple}). Recall that in our notations $\psi =
\rho^{1/2}/(m f_a)$, where $\rho$ is the mass density of axions.

Let us find out when the parametric resonance occurs in axion stars. In
Appendix~\ref{sec:comp-with-bose} we compute $D_{\infty}$
along the line passing through the star center, see
Fig.~\ref{fig:bose_star}. We consider two cases. First, if
self-interactions of axions inside the star are negligible,
Eq.~(\ref{Sinf}) reads,
\begin{equation}
  \label{eq:12}
  M_{s} \geq M_{s,\,0} = 7.66 \, \frac{M_{pl}}{m g_{a\gamma\gamma}}\;,  \qquad\qquad g_4
  \approx 0\;,
\end{equation}
where we restored $g_{a\gamma \gamma} = 2^{3/2}g'/f_a$. This condition is
applicable in the axion-like models with $g_4=0$ or at $M_s \ll
M_{cr}$. In these cases heavier stars are better for the resonance.

Second, if attractive self-interactions are present, the mass  of the axion star is bounded from above, $M_s <
M_{cr}$. Using the profile of the critical star in Eq.~(\ref{Sinf}), we obtain condition 
\begin{equation}
  \label{eq:13}
  g_{a\gamma\gamma} > g_{a\gamma\gamma,\, 0}  \equiv 0.52 \;
  \frac{g_{4}}{f_a}\;, \qquad M_{s} = M_{cr}\;,
\end{equation}
cf.~\cite{Hertzberg:2018zte, Patras2018}.
If this inequality is broken, parametric resonance does not develop
in stable axion stars at all. 

\begin{figure}
  \centerline{\includegraphics{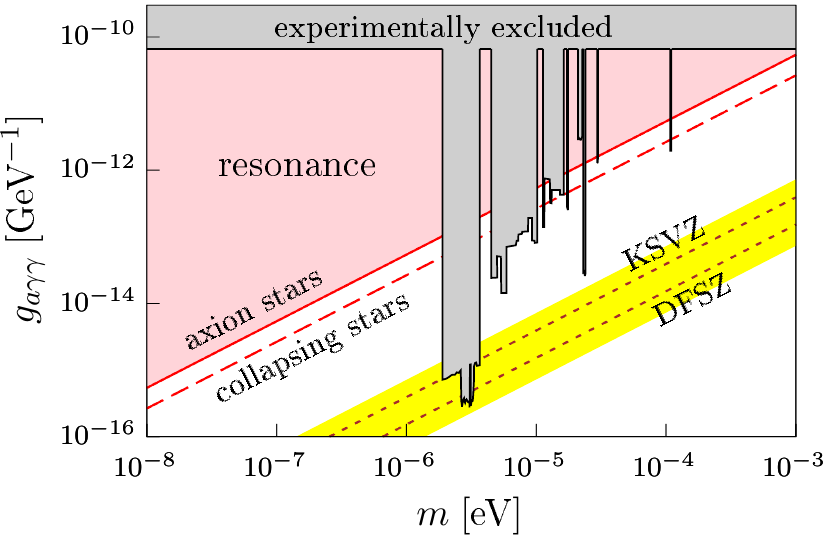}}
  \caption{
    Masses and couplings of QCD axions needed for the Bose stars to
    develop parametric resonance (triangular shaded region above
    the solid line). The respective region for collapsing stars is above the
    dashed line.
    \label{fig:exclusion}}
\end{figure}

For the parameters of QCD axion listed in Sec.~\ref{sec:BS},  the
inequality \eqref{eq:13} gives the shaded region in
Fig.~\ref{fig:exclusion} marked ``resonance.'' 
Notably, the  benchmark values~\cite{diCortona:2015ldu} of
axion-photon coupling (KSVZ-DFSZ band  in Fig.~\ref{fig:exclusion})
are short by two orders of magnitude from igniting the resonance even
in the critical star~\cite{Tkachev:2014dpa,Patras2018,Patras2018}\cite{Wang:2020zur}. On  
the other hand, $g_{a\gamma\gamma}$ is model dependent, with
the only constraint $g_{a\gamma\gamma} < f_a^{-1}$ coming from strong
coupling in simple
models~\cite{Hook:2018dlk,DiLuzio:2020wdo}. Thus, even these simple
  models can satisfy (\ref{eq:13}) within the 
trustworthy parameter range. More elaborated (clockwork-inspired) QCD axion
models~\cite{Agrawal:2017cmd} do not have these limitations and 
easily meet (\ref{eq:13}).

Alternatively, the self-coupling of the axion-like particles
can be arbitrarily small. Condition (\ref{eq:12}) is then satisfied just
for a sufficiently heavy star.

\subsection{Linear exponential growth}
\label{sec:line-expon-growth}
Let us find out how the resonance progresses. One does not expect it to
turn immediately into an exponential catastrophe with $\mu \sim O(L^{-1})$,
like the infinite-volume intuition might suggest,
cf.~Eq.~(\ref{muRsimple}). Rather, the electromagnetic field starts
growing with parametrically small exponent $\mu \ll L^{-1}$
immediately after the condition~(\ref{Sinf}) is met by
the nonrelativistic evolution of axions.  Initial values for this growth
are  tiny. They can be provided by the ambient radiation in
astrophysical setup or, universally, by quantum
fluctuations considered in Appendix~\ref{sec:quantum-start}. In any
case this initial stage proceeds linearly with no backreaction on axions.

We compute the growth exponent by solving the
eigenvalue problem~(\ref{sys})  perturbatively at small 
$\mu$, like in quantum mechanics\footnote{Unlike in quantum mechanics,
  the operator in Eqs.~\eqref{sys} is symplectic, not
  Hermitean.}. To this end we assume that the background $\psi(t,\,
{\bf x})$ did not evolve much from the 
point $\psi_0({\bf x})\equiv\psi(t_0,\, {\bf x})$ when the
condition~\eqref{Sinf} was met, and the resonance mode is close to the 
solution~\eqref{sol}. Calculation in
Appendix~\ref{sec:prop-sympl-oper} gives,
\begin{equation}
  \label{eq:21}
  \mu =  \frac{D_{\infty} - \pi/2}{ \int dz
    \, \sin [2D_0(z)]}\;,
\end{equation}
Here $D_0(z)$ is evaluated using $\psi_0({\bf x})$, a configuration at
the rim of parametric instability, while $D_{\infty}$ uses
$\psi$ in Eq.~\eqref{Sinf}.  Note that application of
Eq.~(\ref{eq:21}) essentially depends on nonrelativistic mechanism  
leading to resonance and providing $D_{\infty} - \pi/2 = O(\psi - \psi_0)$. 

Expression~(\ref{eq:21}) confirms that $\mu$ is indeed
parametrically small and yet, large enough for the  adiabatic 
regime~\eqref{eq:4} to take place. Generically, ${\psi - \psi_0  \sim 
(t_1 - t_0) \,\partial_t \psi}$,   where $t_1 - t_0 \sim \Lambda /\mu$ is
the time from ignition of the resonance to the moment $t_1$ when
the backreaction starts; ${\Lambda \sim \log[C^{\pm}(t_1)
    /C^{\pm}(t_0)] \sim 10^{2}}$ is a large logarithm.  Then
the nonrelativistic scaling~\eqref{eikonal}, \eqref{nonrel} and
Eq.~\eqref{eq:21} give ${\mu \sim \lambda^{-1} (\Lambda
/m\lambda)^{1/2}}$, where we also recalled that the resonance
condition~\eqref{Sinf} is marginally satisfied. Thus,
$$
(m\lambda^2)^{-1} \ll  \mu \ll \lambda^{-1}\;,
$$
i.e.\ the electromagnetic fields evolve faster than the axion
background but slower than the light-crossing time~${L^{-1} \sim
\lambda^{-1}}$.  

Applying Eq.~\eqref{eq:21} to the stationary axion star with $g_4 \approx
0$, we get
\begin{equation}
  \label{eq:44}
  \mu = 0.197\; \frac{m^2}{M_{pl}^2} \, (M_s - M_{s,\, 0})\;,
\end{equation}
where Appendix~\ref{sec:comp-with-bose} was consulted  and $M_{s,\,
  0}$ is given in Eq.~\eqref{eq:12}. Using this expression, one
obtains $\mu \sim 10^{2}\; \mbox{s}^{-1}$ for\footnote{Although we
    use this reference value in all estimates, it is worth stressing
    that presently the mass of the dark matter QCD axion is under debate, 
cf.~\cite{Klaer:2017ond,Buschmann:2019icd}
and~\cite{Gorghetto:2018myk}.} $m=26\, 
\mu\mbox{eV}$~\cite{Klaer:2017ond}
and $\delta M_s \sim 10^{-13}\, M_{\odot}$. Thus, duration of the
linear stage in QCD axion stars is one second or longer. 

\begin{figure}
    \unitlength=1mm
  \begin{picture}(85,51.5)
    \put(0,0){\includegraphics{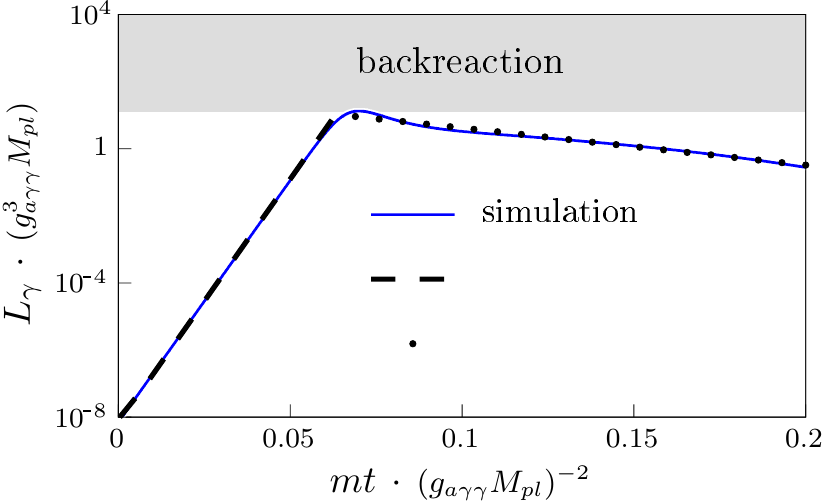}}
    \put(48.5,15.5){Eq.~\eqref{eq:21}}
    \put(48.5,22){Eqs.~\eqref{sys}}
  \end{picture}
  \caption{Luminosity $L_{\gamma}(t) = r^2 \int d\Omega \,
    \boldsymbol{n}_r \, [\boldsymbol{E} \times \boldsymbol{H}]$ of
    axion star with $M_s \approx 1.04 \, 
      M_{s,\, 0}$ during parametric resonance.
      Results of full numerical simulation (solid line) show initial
      growth coinciding with $L_{\gamma} \propto  \exp(2\mu t)$, where
      $\mu$ is given by Eqs.~\eqref{sys} (dashed line). Backreaction is   
       important in  the grey region~\eqref{eq:45}. Late-time
    decay also proceeds exponentially with $\mu$ given by
    Eq.~(\ref{eq:21}) (points) or  Eqs.~\eqref{sys}. 
    Universal units of flux and time are chosen in
    Appendix~\ref{sec:comp-with-bose}.\label{fig:linear}}
\end{figure}

To confirm the above perturbative results, we numerically
solve the system of coupled  relativistic equations~\eqref{equation},
\eqref{eq:5} for the electromagnetic field and axions at $g_4=0$,
see Appendix~\ref{sec:full-relat-simul} for details.
Our simulation starts with the axion star of mass $M_s$ and tiny
electromagnetic amplitudes  representing quantum bath of spontaneous
photons. If the mass of the axion star exceeds $M_{s,\, 0}$, the
exponential growth of amplitudes starts, see the left part of 
Fig.~\ref{fig:linear}. The exponent of this growth coincides with
the one given by Eqs.~\eqref{sys} (dashed line), and within the
expected precision interval of ${\delta\mu/\mu \sim  (M_s - M_{s,\,
  0})/M_{s,\, 0} \sim 4\%}$~--- with Eq.~(\ref{eq:44}).

\begin{figure}
  \unitlength=1mm
  \begin{picture}(85,51.5)
    \put(0,0){\includegraphics{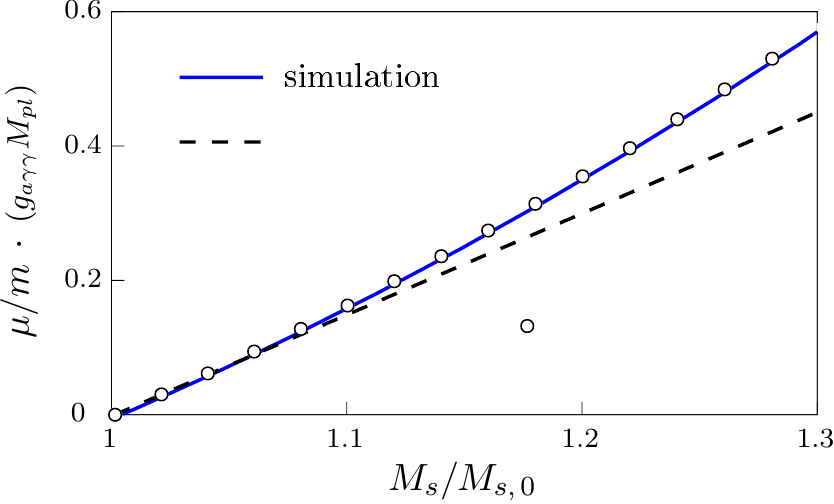}}
    \put(28.8,36){\small Eq.~\eqref{eq:44}}
    \put(57,17.2){\small Eqs.~\eqref{sys}}
  \end{picture}
\caption{The growth exponent $\mu$ as a function of the axion star
  mass $M_s$. Exact   numerical result (solid line) is compared to
  Eq.~(\ref{eq:44}) (dashed) and numerical solution to the non-relativistic problem~\eqref{sys} (points). 
  Units are explained in
  Appendix~\ref{sec:comp-with-bose}.} 
 \label{fig:mu1}  
\end{figure}

In Fig.~\ref{fig:mu1} we show dependence of the exponent $\mu$ on the
axion star mass $M_s$. First, performing full simulations with different stars, we
extract $\mu$ from the exponentially growing flux. This result 
is shown by the solid line.  In the limit
$M_{s}\to M_{s,\, 0}$ it coincides with Eq.~(\ref{eq:44}) (dashed
line), as it should. Second, solving the nonrelativistic
equations~\eqref{sys} numerically, we obtain points in
Fig.~\ref{fig:mu1} which give correct exponent for the arbitrary mass. 

\subsection{Glowing axion stars}
\label{sec:glowing-axion-stars-1}
When the electromagnetic amplitudes in Fig.~\ref{fig:linear}
  become large, the backreaction appears, and the resonant flux
  immediately starts to fall off. Indeed, backreaction burns axions
  diluting their density, and $\mathrm{Re}\, \mu$ in Eq.~\eqref{eq:21}
  decreases to negative values. At this point a long-living
  quasi-stationary level of the electromagnetic field is 
formed. Indeed, at small $\mu <  0$ the resonance mode turns into
  an exponentially growing at $z\to \pm\infty$ solution to Eqs.~\eqref{sys},
\begin{equation}
  \label{eq:54}
  c_x^+ = A\mathrm{e}^{\mu z} \cos D(z)\;,\quad c_y^- = -i A
  \mathrm{e}^{-\mu z} \sin D(z)\;,
\end{equation}
and this is a correct behavior for the quasi-stationary
wave function~\cite{LL3}. Inserting the late-time axion 
configuration   from our full simulation into Eq.~(\ref{eq:21}),
we reproduce the exponential falloff of the flux, see the
dots in Fig.~\ref{fig:linear}. Thus, the solution (\ref{eq:54}),
\eqref{eq:21} remains approximately valid during the entire evolution,
with the only unknown part related to dilution of axions in Eq.~\eqref{eq:8}.

The backreaction switches on when the last term in
  Eq.~\eqref{eq:8} becomes comparable to the others. Using, in
  addition, Eq.~\eqref{Sinf}, we find a condition for the maximal
    flux at the linear stage of resonance, 
    \begin{equation}
  \label{eq:45}
  F_{\gamma} \sim m^2 |C^{\pm}|^2 \lesssim
  \frac{\rho}{m\lambda}\;.
\end{equation}
Here $\lambda$ is the characteristic length scale of axions and and
  $\rho$ is their mass density. In dynamical situations $F_{\gamma,\,
    out}$ is compared to the axion flux $v\rho$ with $v \sim
  (m\lambda)^{-1}$. Notably,
$F_{\gamma} \ll \rho$ when the backreaction starts.
Figure~\ref{fig:linear} demonstrates grey   region where
Eq.~(\ref{eq:45}) is violated.

Let us reconsider the solution~(\ref{sol}), (\ref{S}) with $\mu=0$, to
  describe the regime where the backreaction stops the resonance. The
amplitudes $C_{\alpha}^{\pm}$ of this solution are constant at infinity, 
\begin{equation}
  \label{eq:7}
  C_x^+\big|_{z\to -\infty} = A\;, \qquad \qquad C_{y}^{-}\big|_{z\to
    +\infty} = -iA\;,
\end{equation}
see also Eq.~\eqref{bound}. Thus, the solution describes stationary
flux of photons ${F_{\gamma,\, out}^{\pm}  = \pm m^2 |A|^2}$ from
decaying axions, where for simplicity here and below we assume
equipartition $C_y^+=C_x^+$  and $C_x^- = -C_y^-$. 

Computing the flux~\eqref{eq:9} of produced photons we find $F_{a\to
  \gamma \gamma} = 2m^2 |A|^2 = 2|F_{\gamma,\, out}^{\pm}|$. This means that
the solution~\eqref{sol} duly brings all energy of decaying axions to
infinity.  Energy  conservation law~(\ref{eq:14}) then takes the form,
\begin{equation}
  \label{eq:10}
  \partial_t M = J_{in}   - 2 \int dx dy \, |F_{\gamma,\, out}| \;.
\end{equation}
Even if an arbitrary large constant stream $J_{in}$ of axions is feeded into
the system, the resonance works in the equilibrium regime with
$\partial_t M = 0$ and $\mu= 0$. All arriving axions in this case are
converted into radiation. To break this situation, one needs a
very special mechanism, e.g. the axion star collapse in
  Sec.~\ref{sec:collapse}.

Note that the above stationary situation is stable. Indeed,
perturbing $M$ and $F_{\gamma,\, out}$ away from their equilibrium
values one obtains ${\partial_t \delta M = -2\delta F_{\gamma,\,
  out}}$ due to energy conservation~-- larger flux decreases
the mass. Besides, Eq.~\eqref{eq:21} gives ${\partial_t \delta
  F_{\gamma,\, out} =2\mu F_{\gamma,\, out} \propto \delta 
  M}$ i.e.\ smaller mass weakens the flux. Together, these equations
describe harmonic oscillations around the equilibrium.  In the simplest
uniform model the frequency is ${\Omega   = g_{a\gamma \gamma}
  (F_{in}/8)^{1/2}}$, where $F_{in} = J_{in}/\int dx dy$ is the flux
of axions arriving into the resonance region.  Thus, the resonant
  radioflux $F_{\gamma,\, out}$ may pulsate due to axion-photon
  oscillations. This effect, however, should be strongly dumped due to
  energy dissipation between the modes of the axion field.

In the particular daydream scenario where the Universe is full of
axion stars reaching the condition (\ref{eq:12}) during growth, no
spectacular explosion-like radio events are expected to 
appear in the sky. Most of the axion stars would exist in the
quasi-stationary  regime with ${D_\infty=\pi/2}$, converting all
condensing axions into the radiobackground of frequency
$\omega_\gamma \approx m/2$.

Nevertheless, the latter emission may be observable, even if the
condensation timescale is comparable to the age of the Universe. To
get a feeling of numbers, let us assume that a grown-up star with
$D_{\infty}=\pi/2$  lives  100~pc away from us. Take $m = 26\,\mu\mbox{eV}$
and $M_s \sim 10^{-13} \, M_{\odot}$, the typical values for the QCD
axions. Then the condensation rate onto the star is roughly $10^{-13}
\, M_{\odot}$ per the Universe age. All of condensing axions will be
converted into radiation in the narrow band around $\omega_\gamma \sim
2$ GHz. Even for poor spectral resolution $\delta \omega / \omega \sim
10^{-3}$ one gets spectral flux of order $10^{-2}\, \mbox{Jy}$, which
is detectable. 

When reliable predictions for the abundance of Bose stars and their
growth rates appear, similar calculations  may be used to constrain
the respective scenarios.

\subsection{Amplification of ambient radio}
\label{sec:light-amplification}
Now, we embed the axion stars into astrophysical background of
radiophotons. Namely, suppose an external radiowave of frequency
$\omega_\gamma$ travels through the underdense axion medium which is
safely away from the resonance. The wave will stimulate decay of
axions, so its flux will be amplified in a narrow  spectral window
around ${\omega_\gamma =m/2}$.

This stationary setup  is described by our equations~(\ref{sys})
with $\mu = i(\omega_\gamma - m/2)$ and new boundary conditions,
\begin{equation}
  \label{eq:23}
  \left. c^+_\alpha \right |_{z \to +\infty} = A_0 \;, \qquad \left. c^-_\alpha \right|_{z \to -\infty}
  = 0\; ,
\end{equation}
where equipartition is again assumed and $A_0$ is related to the
incoming electromagnetic flux ${F_{\gamma,\, in} = -m^2 A_0^2}$.

To find the height of the spectral line in this case, we solve
equations at $\omega_\gamma = m/2$ ($\mu=0$). The solution is given by
Eq.~(\ref{sol}) with $A = A_0/\cos D_{\infty}$. The outgoing flux is therefore
\begin{equation}
  \label{eq:24}
  F_{\gamma,\, out} = F_{\gamma ,\, in}/ \cos^2 D_{\infty}\;,
\end{equation}
see also~(\ref{Sinf}). Thus, at small $D_{\infty}$ the extra flux from
axions is weak, $\Delta F = D_{\infty}^2 F_{\gamma ,\, in}$. 
It grows to infinity, however, at $D_{\infty} \to \pi/2$ when the
resonance is about to appear.

For the critical QCD axion stars with $D_{\infty} \ll 1$, 
$$
\Delta F \approx \frac{\pi^2}{4}\;
\frac{g_{a\gamma\gamma}^2}{g_{a\gamma\gamma,\, 0}^2} \; F_{\gamma ,\,
  in}\;,
$$ 
cf.~Eq.~\eqref{eq:13}. In the benchmark KSVZ model
with ${g_{a\gamma\gamma} = 1.92\, \alpha_{em}/(2\pi f_a)}$ this gives $
\Delta F \approx 1.3 \cdot 10^{-4} F_{\gamma ,\, in} $. 
Thus, even underdense axion stars in conservative models shine like
tiny dots on the sky giving narrow spectral lines in excess of smooth
astrophysical background, cf.~\cite{Caputo:2018vmy}.

Let us argue that the Bose stars with ${D_{\infty}\ll 1}$ are better
radioamplifyers than diffuse axions. The latter are described
by kinetic theory~\cite{Tkachev:1987cd, Caputo:2018vmy} which gives
extra amplification ${\Delta F \sim
  g_{a\gamma \gamma}^2 \rho L \lambda\,  F_{\gamma,\, in}}$ from diffuse cloud of size
$L$ and correlation length $\lambda$. We will rederive this expression
  in Sec.~\ref{sec:diffuse-axions} using Eqs.~\eqref{sys}. 
  At  $\lambda \sim L \sim R_s$ it reproduces small-$D_{\infty}$
  result for the axion stars.
  One finds that compact objects give larger amplification, indeed.
  First, if the total mass $M$ is fixed, the product
$\rho L \sim M/L^2$ is larger for smaller $L$. Second, the  
wavelengths $\lambda \sim 10^{3}\, m^{-1}$ of diffuse axions in  
the Galaxy are much smaller than the radii of axion stars.

Let $Q$ be the fraction of dark matter in the axion stars. Stimulated
emission from these objects in our Galaxy is suppressed by the
tiny geometric factor  $R_s^2 / L^2$, where $L\sim \mbox{kpc}$, as
compared to diffuse axions. However, multiplying it by the above
boost factor, we find  
${\Delta F_{\mathrm{stars}}/\Delta F_{\mathrm{diffuse}} \sim Qmv
  R_s}$, where $v\sim 10^{-3}$ is the velocity of diffuse axions. For
critical QCD axion stars and $m=26\, \mu\mbox{eV}$ this ratio equals
$QvM_{pl}/ f_a \sim 10^{4} \, Q$, so the stars give larger
stimulated flux at $Q \gtrsim 10^{-4}$. 

Finally, in the scenario with enhanced axion-photon coupling our Universe
may be full of quasi-stationary axion stars with $D_{\infty} \lesssim 
\pi/2$. A radiowave passing through one of these objects burns
essential fraction of its axions producing a powerful flash of
radio-emission\footnote{In Eq.~(\ref{eq:24}) we ignored backreaction of
  photons on axions which may be relevant in this case.}. This effect
can be used to constrain some arrogant models. 

\subsection{Radio-portrait of an axion star}
\label{sec:prof-reson-mode}
In generic resonating axion cloud there exists one,  at most several
directions where the condition~\eqref{Sinf} is satisfied. Parametric
emission forms narrow beams pointing in these
directions. But the Bose stars are spherical, with all diameters
giving the same $D_{\infty}$.  The question is, what is the
distribution of the resonant flux in angular harmonics.

\begin{figure}
  \centerline{\includegraphics{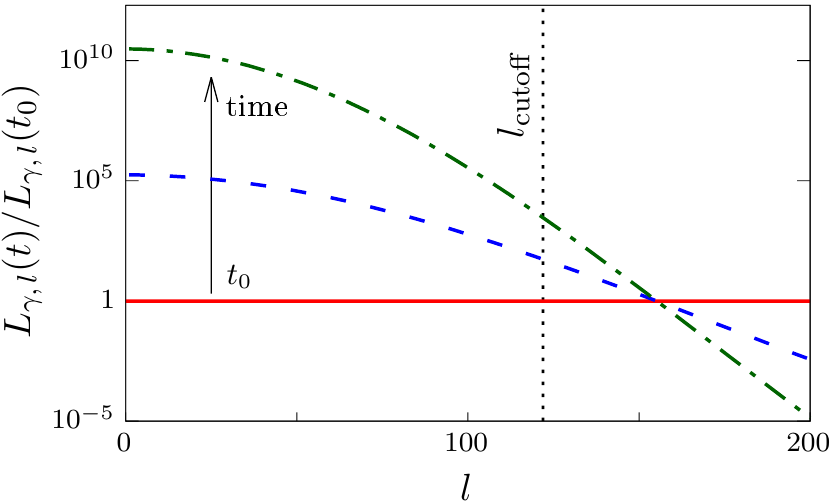}}
  \caption{Luminosity distribution over angular harmonics
    $L_{\gamma,\, l}(t)$. We consider resonant emission
    from the stationary Bose star with ${M_s = 1.36\, M_{s,\,
        0}}$, ${mR_s \sim M_{pl}^2 / (mM_s) \approx 115}$,
    and $g_4=0$.  Lines are the fixed-time sections of 
    luminosity in full numerical
    simulation.\label{fig:angular_distribution}}
\end{figure}

In Appendix~\ref{sec:spherical-symmetry} we perform spherical
decomposition of the electromagnetic field inside  a  Bose star. We find
the same leading-order equations~\eqref{sys} in every angular
sector
$(l,\, m')$, with dependence on $l$ emerging as an $O(mR_s)^{-1}$ correction to
the spatial derivatives
\begin{equation}
  \label{eq:46}
    \partial_z \to \partial_z + \frac{i l(l+1)}{mz^2}\;,
\end{equation}
where $z=\pm r$, cf.\ Eq.~\eqref{eq:35}. In fact, even this correction can
be absorbed by the singular redefinition ${c_{i}^\pm \to c_{i}^{\pm}  \exp[i   
  l(l+1)/mr]}$ of the electromagnetic amplitudes. Then the effect
of angular quantum number is parametrically weaker than $(mR_s)^{-1}$, with
leading contribution coming from a small vicinity of~${r=0}$. We conclude
that spherical modes with essentially different $l$ 
satisfy almost the same equations inside the star and grow at close 
rates $\mu_l\approx \mu$. 

Our numerical simulation confirms this expectation, see
Fig.~\ref{fig:angular_distribution}. Namely, the numerical data
suggest heuristic expression\footnote{We do think that
  Eq.~\eqref{eq:62} can be derived perturbatively. However, this
  calculation goes beyond the scope of this paper.},
\begin{equation}
  \label{eq:62}
  \mu_l - \mu \approx -0.034 \, m \, l(l+1)\,\frac{M_{s}}{M_{s,0}} \, \left(
  \frac{mM_s}{M_{pl}^2}\right)^3 \;,
\end{equation}
where $\mu$ is approximately given by Eq.~\eqref{eq:44}. Thus,
dependence on $l$ is indeed an $O(mR_s)^{-2}$ correction, see
Appendix~\ref{sec:comp-with-bose}.

\begin{figure}
  \centerline{\includegraphics{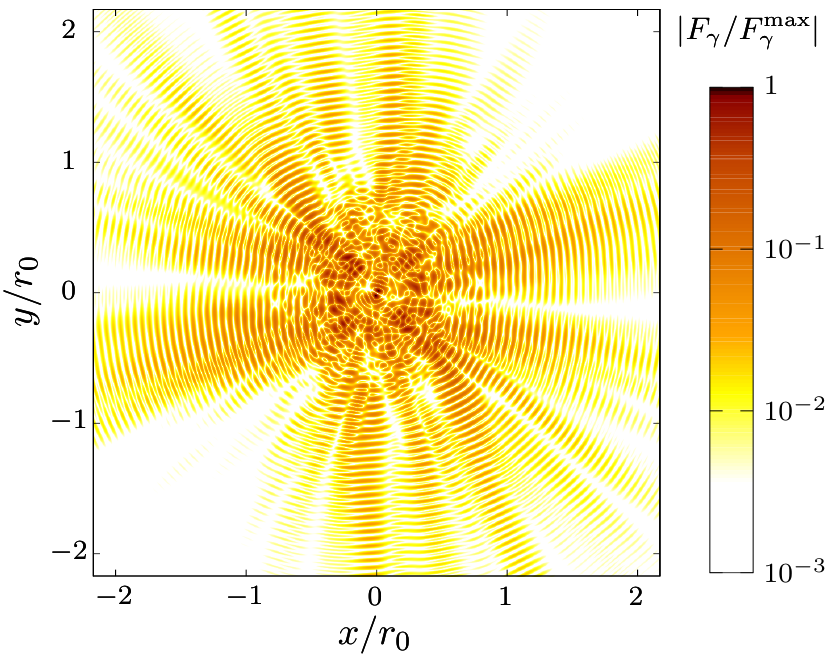}}
  \caption{Electromagnetic flux $F_{\gamma}\equiv \boldsymbol{n}_r
    [\boldsymbol{E}\times \boldsymbol{H}]$  inside the resonating star
    from Fig.~\ref{fig:angular_distribution}; $r_0 = M_{pl}^2/(M_s
    m^2)$. Interference between the 
    waves moving in the $+r$ and $-r$ directions is clearly seen. The simulation
    uses random initial data to mimic quantum fluctuations in
    the electromagnetic vacuum, see
    Appendices~\ref{sec:quantum-start},~\ref{sec:full-relat-simul} for 
    details.\label{fig:chaos}}
\end{figure}

Now, it is explicit that all modes with
$$
l \lesssim l_{\mathrm{cutoff}} \approx 2.4 \, \left(
\frac{M_s}{M_{s,\, 0}} - 1\right)^{1/2} \, \frac{M_{s,\, 0}
  M_{pl}^2}{m M_s^2} \sim mR_s
$$
grow simultaneously in resonance, see the vertical dotted
line\footnote{The line is $30\%$ off because we used
    Eq.~(\ref{eq:44}) which has accuracy ${(M_s - M_{s,\, 0})/M_{s,\,
        0} \sim 0.4}$. For better precision
    one has to compute $\mu$ in Eqs.~\eqref{sys} numerically  and obtain
    $l_{\mathrm{cutoff}}$ from Eq.~(\ref{eq:62}) at~${\mu_{l}\approx 0}$.} in   
Fig.~\ref{fig:angular_distribution}. If the instability starts from
random quantum fluctuations, it produces chaotic angular distribution
in Fig.~\ref{fig:chaos} with typical angular size
$l_{\mathrm{cutoff}}^{-1}$. If the instability starts due to
ambient radiowave, the cutoff sets typical width of the resonance
beam.

\subsection{Two axion stars}
\label{sec:phases-are-important}

Suppose two Bose stars came close to each other with negligible
relative velocity. Together, their profiles may satisfy the resonance
condition even if the individual  stars are far away from it. Then
strong and efficient parametric resonance may develop in this
system~\cite{Hertzberg:2018zte}.

We describe this case considering the background
\begin{equation}
  \label{eq:27}
  \psi = \psi_{s}({\bf x})\, \mathrm{e}^{i\theta_s} +
  \psi_{s}'({\bf x})\, \mathrm{e}^{i\theta_s'}\;,
\end{equation}
of well separated static Bose stars $\psi_s$ and $\psi_s'$ centered at
$z=0$ and $z=L$, respectively. In Eq.~(\ref{eq:27}) we explicitly
introduced complex phases of stars $\theta_s$ and $\theta_s'$.

Equations~\eqref{sys} can be solved analytically in the limit when the
interstar distance is much larger than their sizes, $L \gg R_s$. In
this case $\mu \sim O(L)^{-1}$ corresponds to the inverse
light-crossing time between the stars. Outside every star i.e.\ at 
$z \ll L$ and at $z \gg 0$, we obtain
\begin{align}
  \label{eq:20}
  & \left.\begin{array}{l}
      \displaystyle c_x^+ = A\, \mathrm{e}^{\mu z}\,  \cos D(z)\\[2px]
      c_y^- = -i A\, \mathrm{e}^{i\theta_s - \mu z} \sin D(z)
      \end{array}\right\} \; \mbox{outside $\psi_s'$},\\
  \label{eq:25}
  & \notag\left.\begin{array}{l}
      c_x^+ = A' \,\mathrm{e}^{\mu (z-L)}\,  \sin[D'_{\infty} - D'(z)]\\[2px]
      c_y^- = -i A'\, \mathrm{e}^{i\theta_s' - \mu (z-L)} \cos
      [D'_{\infty} - D'(z)]
    \end{array}\right\} \; \mbox{outside $\psi_s$},
\end{align}
where $D$ and $D'$ are computed using $\psi_s$ and $\psi_s'$ in
Eqs.~\eqref{S}, \eqref{Sinf}.   Indeed, 
expressions~(\ref{eq:20}) satisfy the boundary value
problem inside the left and right stars with 
$O(\mu)$ precision, and both of them give correct solution
between the stars.  Gluing $c_x^+$ and $c_y^-$ in the latter region, one finds
$A = A' \mathrm{e}^{-\mu L} \sin D'_\infty/\cos D_\infty$ and 
\begin{equation}
  \label{eq:26}
  \mu = \frac{1}{2L} \left[ i \theta_s - i\theta_s' + \ln \; \frac{\sin
    D_{\infty} \, \sin D'_{\infty}}{\cos D_{\infty} \, \cos D'_{+\infty}} \right]\;,
\end{equation}
which confirms that $\mu \sim O(L)^{-1}$.

Expression (\ref{eq:26}) deserves discussion. First, the two-star
system hosts parametric resonance if
$\mathrm{Re}\, \mu \geq 0$ or 
$D_{\infty}+D'_{\infty} \geq \pi/2$. This condition reproduces the 
naive criterion~\eqref{Sinf} with $\psi \to |\psi|$. Second, the
resonance develops at a very slow rate $\mu \sim L^{-1}$ which
is nevertheless much faster than the evolution of $\psi$ if $\mu
\gg \omega_s$ or $mR_s^2 \gg L$.

Third and importantly, left- and right-moving parametric waves have
slightly different frequencies ${\omega_\gamma   = m/2 \pm
  \mathrm{Im}\, \mu}$, where $\mathrm{Im}\, {\mu} = (\theta_s -
  \theta_s')/2L$, cf.\ Eq.~\eqref{eq:4}. This splitting is a benchmark
effect of incoherent axions. Technically, it 
appears because the phases of the resonant amplitudes are
locally related to the phase of the axion field,  
\begin{equation}
  \label{eq:29}
  \arg\, c_x^+ \approx \arg c_y^- - \mathrm{arg}\, \psi + \pi/2\;.
\end{equation}
Indeed, all coefficients in Eqs.~(\ref{sys}) become real after
substitution $c_y^- \to ic_y^- \exp (i \, \mathrm{arg}\, \psi)$ with
corrections suppressed by $\partial_z \, \mathrm{arg}\, \psi$; hence
(\ref{eq:29}). In the above setup 
with two axion stars the shifts of emission frequencies
ensure Eq.~(\ref{eq:29}) inside each star at $z\approx 0$ and~$L$.

Notably, one does expect formation of gravitationally bound groups
  of Bose stars in the
  QCD axion cosmology. Indeed, in the post-inflationary scenario these
  objects emerge in the centers of miniclusters which are organized in
  chains and hierarchically bound
  structures~\cite{Vaquero:2018tib,Buschmann:2019icd,Eggemeier:2019khm}.
Once several stars within one group align with small relative
velocities $v \ll (mL)^{-1}$, condition~\eqref{Sinf} may be
satisfied and the parametric explosion  follows. The spread of the
produced spectrum will be $\delta \omega_\gamma / \omega_\gamma  \sim
L^{-1}$ due to random phases of the stars, even if their velocities
are negligibly small. 

\section{Diffuse axions}
\label{sec:diffuse-axions}

Our eikonal system~\eqref{sys} is a microscopic Maxwell's
equation in disguise. It is valid for general axion backgrounds
including virialized distributions in the galaxy cores and
axion miniclusters. In the latter cases, however, kinetic
approach is simpler.

In this Section we study parametric radio-amplification in a
cloud of random classical waves representing incoherent or partially
coherent axions. We fix correlators
\begin{equation}
  \label{eq:49}
  \langle \psi \rangle = 0 \;,\;\; 
  \langle \psi^*(z) \psi(z') \rangle = \rho\;  C(z - z')/(mf_a)^2\;,
\end{equation}
where $ \rho$ is density, $C(0)=1$, and the correlation length is
$\lambda = \int dy \, C(y)$.  

\begin{figure}
  \centerline{\includegraphics{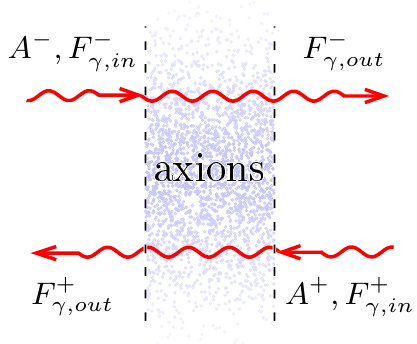}
    \includegraphics{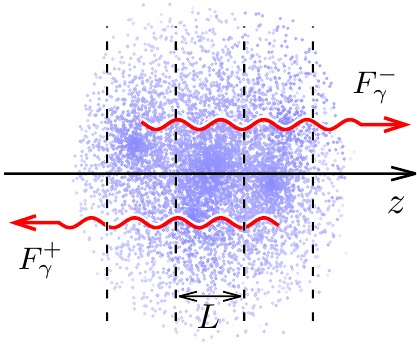}}
  \centerline{(a) \hspace{4cm} (b)}
  \caption{{ (a) Radiowaves going through a small region with 
    axions. (b) Two resonant radiofluxes in a large axion 
    cloud.}\label{fig:diffuse}}
\end{figure}

Let us coarse-grain Eqs.~\eqref{sys} to a kinetic equation in 
the stationary case. To this end we consider two radiowaves with 
fixed frequency $\omega_\gamma = m/2$ and amplitudes $A^{\pm}$
traveling back-to-back through a small axion region  in
Fig.~\ref{fig:diffuse}a. This fixes the boundary conditions, 
\begin{equation}
  \label{eq:50}
  \left. c^+_x \right |_{z \to +\infty} = A^+ \;, \qquad
  \left. c^-_y \right|_{z \to -\infty} = A^-\;,
\end{equation}
and the incoming fluxes $F_{\gamma,\, in}^{\pm} = \mp m^2
|A^{\pm}|^2/2$.

We assume that by itself, the axion region is too small to host
a resonance. Then the nonrelativistic equations~\eqref{sys},
\eqref{eq:50} can be solved perturbatively,
\begin{align}
  \notag
  &c_x^+ = A^+ \left[1 + D_{2,\, \infty} - D_2(z) \right] + i A^- \left[
    D_\infty^* - D^*(z)\right]\;,\\
  \label{eq:57}
  &c_y^- = A^- \left[1 + D_{\infty}^* D(z) - D_2^*(z)\right] - iA^+ D(z)\;,
\end{align}
where $D(z)$ is given by Eq.~\eqref{Sinf} and 
\begin{equation}
  \label{eq:58}
  D_2(z) = {g'm} \int_{-\infty}^z dz' \, \psi^*(z') \, D(z')\;.
\end{equation}
We compute the outgoing fluxes by performing ensemble average
via Eq.~\eqref{eq:49},
\begin{equation}
  \notag
  F_{\gamma,\, out}^+ = - \frac{m^2}{2} \langle |c_x^+|^2\rangle_{z\to -\infty}, \;
  F_{\gamma,\, out}^- =   \frac{m^2}{2} \langle  |c_y^-|^2\rangle_{z\to +\infty}.
\end{equation}
The solution~\eqref{eq:57} gives, 
\begin{equation}
  \label{eq:61}
  F_{\gamma,\, out}^{\pm} = F_{\gamma,\,  in}^{\pm} (1  + \mu_\infty' L )  -
  \mu_{\infty}' L \, F_{\gamma,\, in}^{\mp} \;.
\end{equation}
Here $L$ is the size of the axion region and ${\mu_\infty' =
  \langle |D_{\infty}|^2 \rangle/L}$ is the naive growth exponent in the
  infinite axion gas. The latter parameter is explicitly computed by
  assuming that the region is macroscopic, $L\gg \lambda$, and yet, small at the
  scales of~$\rho$, 
\begin{equation}
  \label{eq:63}
  \mu_\infty' = g_{a\gamma\gamma}^2 \rho \lambda / 8 \;,
\end{equation}
where we restored the physical coupling $g_{a\gamma\gamma}$. 

Now, consider large axion cloud. We divide into small regions of width
$L$, see  Fig.~\ref{fig:diffuse}b. Since equation~(\ref{eq:61}) is
valid in every region, we find,
\begin{equation}
  \label{eq:64}
  \partial_z F_{\gamma}^{\pm} = \mu_{\infty}'(z) (F_{\gamma}^- - F_{\gamma}^+)\;,
\end{equation}
where $F_\gamma^{\pm}(z)$ are the fluxes $F_{\gamma,\, in}^{\pm} \approx
F_{\gamma,\, out}^{\pm}$ at the macroscopic position $z$.

Recalling that $F_{\gamma}^{+}$ and $F_{\gamma}^{-}$ travel in $-z$
and $+z$ directions, respectively, one
restores the time derivative in Eq.~(\ref{eq:3}) by changing
\begin{equation}
  \label{eq:3}
  \partial_z F^\pm \to (\partial_z \mp \partial_t) F^\pm\;.
\end{equation}
After that our kinetic equation coincides with the one in
Refs.~\cite{Tkachev:1987cd,Caputo:2018vmy} if one trades the
correlation length $\lambda(z)$ in Eq.~(\ref{eq:63}) for 
the axion velocity ${v \sim (m\lambda)^{-1}}$ or spectral width of
radiowaves $\delta \omega_\gamma \sim \lambda^{-1}$.

Solving Eq.~(\ref{eq:64}) in the stationary case, we find,
\begin{equation}
  \label{eq:65}
  F_{\gamma}^{-}(z) = F_{\gamma}^+(z) + F_{0}  = F_0\int\limits_{-\infty}^{z}
  \mu_{\infty}'(z') \,dz'\;,
\end{equation}
where $F_0$ is the integration constant. Note that this solution
  does not indicate exponential growth of fluxes, unlike the
  time-dependent solutions of Eqs.~\eqref{eq:64}, \eqref{eq:3} 
  behaving like $F^\pm_\gamma \propto \exp(\mu'_\infty t)$ in the
  infinite medium.

Nevertheless, one can use Eq.~(\ref{eq:65}) for waves with
${\omega_{\gamma} = m/2}$ ($\mu=0$) in two important respects. First,
$\mu=0$ when the resonance is about to  appear.  In this case the
ambient fluxes are absent: $F_{\gamma}^{+}(+\infty)=
F_{\gamma}^{-}(-\infty) = 0$, cf.\ Eq.~\eqref{bound}. The
solution~(\ref{eq:65}) satisfies this criterion only at $D_{\infty,\,
  \mathrm{diff}} = 1$, i.e.\ at the boundary of the region
\begin{equation}
  \label{eq:66}
  D_{\infty,\, \mathrm{diff}} \equiv  \frac{g_{a\gamma\gamma}^2}{8}
  \int_{-\infty}^{+\infty} \rho (z) \lambda(z)\, dz \geq 1\;. 
\end{equation}
This inequality gives precise condition for the parametric resonance
in diffuse axions, cf.\ Eq.~\eqref{Sinf}. 

Second, even far away from the parametric instability Eq.~(\ref{eq:65})
predicts amplification of ambient radioflux $F_{\gamma,\, in} =
F^+(+\infty)$ due to decay of axions,
$$
{F_{\gamma,\, out} = F_{\gamma,\, in} \,/ (1 -
  D_{\infty,\, \mathrm{diff}})}\;,
$$
where $F_{\gamma,\, out} = F^+(-\infty)$, cf.\ Sec.~\ref{sec:light-amplification}. 

\section{Moving axions}
\label{sec:moving-axions}
\subsection{Doppler shifts and new resonance condition}
\label{sec:generalities}

We just saw that motion of diffuse axions decreases their
correlation length $\lambda \sim (mv)^{-1}$ and hence suppresses the
resonance, cf.\ Eq.~\eqref{eq:66}. In this Section we study the effect
of moving coherent axions. 

Let us rewrite the system~\eqref{sys} in terms of physical parameters:
axion velocity $v_i(t,\, {\bf x})$ in Eq.~\eqref{eq:6}, and density
$\rho (t,\, {\bf x}) = m^2 f_a^2 |\psi|^2$. To this end we change
variables, 
\begin{equation}
  \label{eq:28}
  c_x^+ = \tilde{c}_x^+ \, \mathrm{e}^{-i\mathrm{arg}\,\psi /2}\;,
  \qquad c_y^- = \tilde{c}_y^- \, \mathrm{e}^{i\mathrm{arg}\,\psi /2}\;.
\end{equation}
Eikonal equations take the form, 
\begin{subequations}
  \label{sysrho}
\begin{align}
  \label{sysrho1}
  &\left(2\mu + im\, v_z\right) \tilde{c}^+_x = 2\partial_z
  \tilde{c}^+_x + i g_{a\gamma\gamma} (\rho/2)^{1/2} \; \tilde{c}^-_y\;, \\
  \label{sysrho2}
  &\left(2\mu + im\, v_z \right) \tilde{c}^-_y = - 2\partial_z \tilde{c}^-_y 
  - ig_{a\gamma\gamma} (\rho/2)^{1/2} \;\tilde{c}^+_x\;.
\end{align}
\end{subequations}
Note that only a projection $v_z$ of the axion velocity to the
resonance axis matters. 

If $v_z$ is constant, one can eliminate it from Eqs.~(\ref{sysrho}) by
changing $\mu \to \mu - imv_z/2$. This is the  Doppler shift of
frequencies ${\omega_{\gamma} = m/2 \pm \mathrm{Im}\,\mu}$ for the left-
and right- moving waves in Eq.~\eqref{plane}. Apart from that,
constant velocities do not affect the resonance at
all. Indeed, one can always transform to the rest frame of axions.

The situation changes if some parts of the axion matter move with respect to
others: ${v_z = v_z(z)}$. Then the axions decaying in various
parts produce photons with different frequencies, and this kills 
Bose amplification of induced decays. Thus, relative velocities 
are the main show-stoppers for the  parametric resonance.  

\begin{figure}
\includegraphics{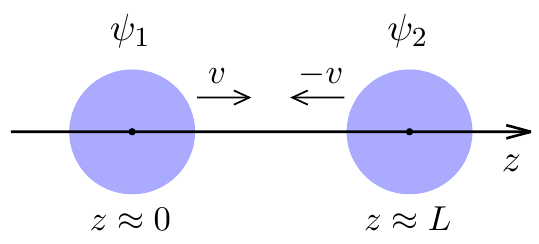}
\caption{Two moving Bose stars.}
\label{fig:two_chunks}
\end{figure}

In the next Section we will demonstrate that only the coherent regions
with relative velocities
\begin{equation}
  \label{eq:33}
  v\lesssim (mR)^{-1}\;,
\end{equation}
can be simultaneously in resonance, where $R$ is the size of
these regions. The above expression is natural. Indeed, $R^{-1}$ is the
momentum spread in the resonance mode. If the Doppler shift $mv$ is
larger, photons produced in different regions are out of resonance. 

\subsection{Two moving axion stars}
\label{sec:two-moving-axion}
\begin{figure}
\includegraphics{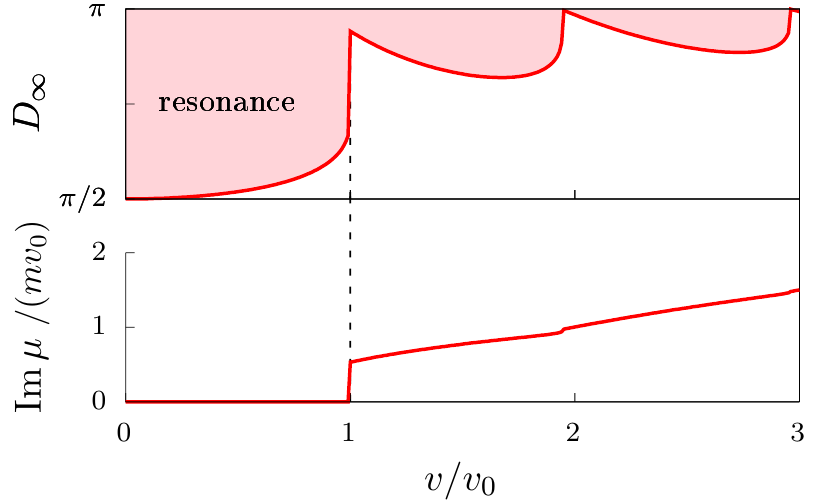}
\caption{Condition for parametric resonance in two moving axion stars
  (top panel) and respective Doppler shift $\mu' = \mathrm{Im}\,\mu$
  (bottom panel).} \label{fig:phase_transition}
\end{figure}

To get a qualitative understanding of relative velocities, we consider two
identical Bose stars approaching each other at a nonrelativistic constant speed $v$, 
$$
\psi = \psi_1(z) \, \mathrm{e}^{imvz} + \psi_{2}(z) \, \mathrm{e}^{-imv(z-L)}\;,
$$
see Fig.~\ref{fig:two_chunks}. For simplicity we will assume that
$\psi_1$ and $\psi_2$ are equal to a constant $\psi_0$ in the regions
$0< z < 2R_s$ and ${L < z < L + 2R_s}$, and they are zero  outside.
We are going to find out whether this configuration develops a
  resonance before the merger i.e.\ when the profiles of the stars
  still do not overlap.

We compute the resonant mode by solving Eqs.~\eqref{sysrho} in 
the regions of constant $\rho$, $v_z$ and  gluing the original
  amplitudes $c_{x,y}^{\pm}$
at $z = 2R_s$ and $z = L$. Then the boundary conditions~\eqref{bound}
give equation for the growth exponent $\mu$. At the border of 
resonance ${\mu= i\mu'}$ becomes imaginary and the equation simplifies, 
\begin{multline}
  \label{eq:30}
  \mathrm{tan}^2(2\kappa_- R_s) \mathrm{tan}^2 (2\kappa_+ R_s) = \left[ 1
    +\frac{(2\mu' - mv)^2}{m^2v_0^2\cos^2
      (2\kappa_- R_s)} \right]\\
 \times \left[ 1 + \frac{(2\mu' + mv)^2}{m^2v_0^2\cos^2 (2\kappa_+ R_s)} \right]\;.
\end{multline}
Here we introduced the relevant velocity scale ${v_0 = 2g'
  \psi_0}$ and notations $4\kappa_{\pm}^2 = m^2
v_0^2 +  (2\mu'\pm mv)^2$. 

At a very naive level, one may use $|\psi|$ instead of $\psi$ in
Eq.~\eqref{Sinf}. Then the resonance is expected at
${D_{\infty} \equiv 4g'm \psi_0 R_s \geq\pi/2}$, where
$D_{\infty}$ sums up contributions from both stars. In truth, 
the solution of Eq.~\eqref{eq:30} exists only in the shaded
region in  Fig.~\ref{fig:phase_transition} (top panel). The Doppler
shift $\mu' = \mu'(v)$ at the boundary of this region is plotted in
the bottom panel. 

One observes sharp first-order phase transition at ${v\approx v_0}$
between the two resonance regimes, see the vertical dashed line in
Fig.~\ref{fig:phase_transition}. At $v< v_0$ the Doppler shift 
is absent, $\mathrm{Im}\, \mu=0$, although the stars have nonzero
velocities. Besides, the naive resonance condition $D_{\infty}\geq \pi/2$ is
approximately valid indicating that the instability develops
simultaneously in both stars.  To the contrary, at $v > v_0$ two
individual stars  host their own resonances,  
with little help from each other. In this case the Doppler shift is
$\mathrm{Im}\, \mu \approx mv/2$ and the resonance condition 
$D_{\infty}/2 > \pi/2$ coincides with that for one star. We
conclude that the two-star resonance occurs only at $v \leq v_0$ or
Eq.~(\ref{eq:33}).  

Note that the phase transition in Fig.~\ref{fig:phase_transition} can
be understood analytically. At large relative velocities ${v\gg v_0}$
at least one of the two brackets in the right-hand side of
Eq.~(\ref{eq:30}) should be small, so the solutions are ${\mu' \approx
  \pm mv/2}$ and $2\kappa_{\pm} R_s  \approx D_{\infty}/2\approx
\pi/2$. This corresponds to resonance in individual stars. At $v\lesssim
v_0$ Eq.~(\ref{eq:30}) with $\mu'=0$ 
takes the form
$$
\cos(4\kappa_{\pm} R_s) = -v^2 /v_0^2\;,
$$
where $\kappa_{\pm} = m(v_0^2 + v^2)^{1/2}/2$. At  ${v \ll v_0}$ we obtain
${D_{\infty} = \pi/2}$, --- a condition for the two-star resonance.
  At $v > v_0$ the above equation in the case $\mu'=0$ does not have
  solutions. 
  
\subsection{Collapsing stars}
\label{sec:collapse}

\begin{figure}
  \hspace{1.3cm}(a)\hspace{3.5cm}(b)\\
  \includegraphics{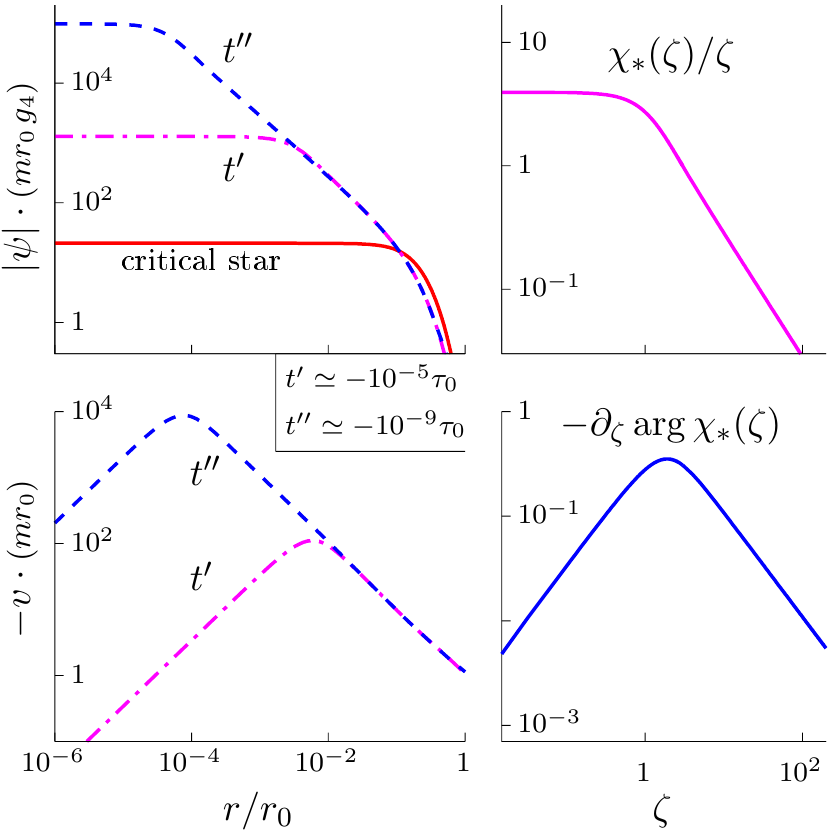}
  \caption{(a) Numerical solution to the Schr\"odinger-Poisson
    system~\eqref{eq:36}, \eqref{eq:40} describing collapse of a
    critical Bose star; the axion velocity is $v = m^{-1}\,
    \partial_r \mathrm{arg}\, \psi$. We use space and time units ${r_0
      = g_4 M_{pl}/(mf_a)}$ and $\tau_0 = mr_0^2$, see
    Appendix~\ref{sec:comp-with-bose}. (b)~Universal 
    self-similar attractor.}
  \label{fig:collapse}
\end{figure}

Now, consider collapse of a critical axion star,  ${M_s = M_{cr}}$, caused
by the attractive self-interaction of axions. During this process the
axions fall into the star center acquiring velocities and making the
density grow, see Fig.~\ref{fig:collapse}a. These two effects suppress
the resonance and support it, respectively.

We are going to study the resonance at the first stage of the
  collapse when the infalling axions are still nonrelativistic and
  their field is weak, $|\psi| \ll 
  1$. In this case the Schr\"odinger-Poisson 
  system~\eqref{eq:36},~\eqref{eq:40} for axions is applicable,
  whereas the electromagnetic field is described by
  Eqs.~\eqref{sys}.

\begin{figure}[ht]
  \begin{tabular}{cc}
    \unitlength=1mm
    \begin{picture}(85,65)
    \put(0,0){\includegraphics{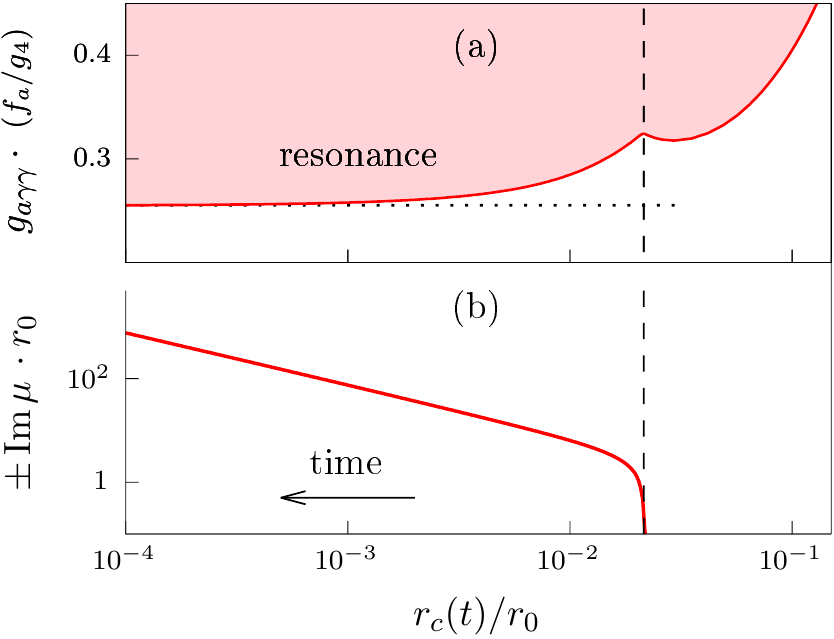}}
    \put(70,43){\small Eq.~\eqref{eq:59}}
  \end{picture}
  \end{tabular}
  \caption{(a) Electromagnetic
    coupling $g_{a\gamma\gamma}$ required  for parametric resonance in
    collapsing critical star at the moment when its core radius  is
    $r_c(t)$. (b) Doppler shifts $\pm\mathrm{Im}\, \mu$ at the moment 
    of ignition. Unit of length is ${r_0 = g_4  M_{pl}/(mf_a)}$.
    \label{fig:gmin}
  }
\end{figure}

To find out how the parametric instability progresses,  we
numerically solve the boundary value problem~\eqref{sys} in the
background $\psi(t,\, r)$ of the collapsing star at every $t$.
We characterize the stage of collapse with the radius ${r=r_c(t)}$
where
the axion field drops by a factor of two from its value  in the
center: ${|\psi(t,\, r_c(t))| = |\psi(t,\, 0)|/2}$.  We will see that 
the region  $r\lesssim r_c$ is  important for the resonance
despite the fact that $r_c(t)$ decreases by orders of magnitude during
collapse. Shaded region Fig.~\ref{fig:gmin}a covers couplings
$g_{a\gamma\gamma}$ required for the resonant solutions of
Eqs.~\eqref{sys} to exist at time $r_c(t)$.
 At the lower boundary of this region $\mathrm{Re}\, \mu=0$; the
respective Doppler shifts $\mathrm{Im}\, \mu$ are presented in
Fig.~\ref{fig:gmin}b.

Since the star is spherically-symmetric,  ${\psi(z) =
  \psi(-z)}$,
the photon modes with complex exponents $\mu$ appear in conjugate
pairs. Indeed, for every solution $\{c_x^+(z),\, c_y^-(z)\}$ of
Eqs.~\eqref{sys} with eigenvalue $\mu$, there exists a solution
$\{[c_y^-(-z)]^*,\, [c_x^+(-z)]^*\}$ with eigenvalue
$\mu^*$. Physically, this means that for every axion there
  exists a diametrically opposite axion with the opposite velocity giving 
Doppler shift $-\mathrm{Im}\, \mu$. Two signs in the ordinate label of
Fig.~\ref{fig:gmin} represent these two solutions. 

In Fig.~\ref{fig:gmin} we again see the first-order phase transition
(vertical dashed line) described in
Sec.~\ref{sec:two-moving-axion}. Indeed, if the resonance  appears
immediately after the collapse begins (large $r_c$), it involves all
slowly-moving axions and develops with $\mathrm{Im}\, \mu=0$. At later
stages of collapse (smaller $r_c$) the resonance can be supported
only by fast axions in the dense star core, hence the Doppler shift
$\mathrm{Im}\, \mu \ne 0$. Importantly and unlike in the previous 
Section, the stage with fast axions is better for resonance, as it
can occur at smaller couplings, cf.\  Figs.~\ref{fig:gmin}a
  and~\ref{fig:phase_transition}. 

We therefore consider resonance in the central core of a collapsing
star. It was shown~\cite{Zakharov12,Levkov:2016rkk} that evolution
of the axion field in this region is described by the universal
self-similar attractor,
  \begin{equation}
    \psi(t, r) = \frac{(-m t )^{-i\omega_*} }{m r g_4} \chi_*
    \left(\zeta\right) \;, \qquad  \zeta= r \, \sqrt{-m/t}\;,
    \label{eq:sscol}
  \end{equation}
where $t<0$, $\omega_* \approx 0.54$ and the function
$\chi_*(\zeta)$ is presented in Fig.~\ref{fig:collapse}b. The core size $r_c
(t) \approx 1.5 \, (-t/m)^{1/2}$ shrinks from the macroscopic values $r_c
\sim R_s$ to $m^{-1}$ during self-similar stage. Without the
  parametric resonance into photons,  relativistic
corrections become relevant~\cite{Levkov:2016rkk} at the end of this
stage $t\gtrsim -m^{-1}$. Simultaneously,  the weak--field
  approximation gets broken and higher-order terms of the axion
  potential \eqref{eq:aPot} become essential. Below we concentrate on
  the   situations when the resonance starts at the nonrelativistic
  stage $t \ll -m^{-1}$.

Substituting Eq.~\eqref{eq:sscol} into the spectral
problem~\eqref{sys} and changing variables $c^\pm = (-mt)^{\pm
  i\omega_*/2}\, \tilde{c}^{\pm}(\zeta)$, we arrive to time-independent
spectral problem
  \begin{subequations}
  \label{eq:56}
  \begin{align}
    \label{eq:53}
    &\tilde{\mu} \tilde{c}_x^+ = \partial_\zeta \tilde{c}_x^+ +
    \frac{ig'}{g_4} \, \frac{[\chi_*(\zeta)]^*}{\zeta} \, \tilde{c}_y^-\;,\\
    &\tilde{\mu} \tilde{c}_y^- = -\partial_\zeta \tilde{c}_y^- - 
    \frac{ig'}{g_4} \, \frac{\chi_*(\zeta)}{\zeta} \, \tilde{c}_x^+\;,
  \end{align}
  \end{subequations}
  which involves only one combination of parameters $g'/g_4$. We also
introduced
\begin{equation}
  \label{eq:55}
  \mu = \tilde{\mu} \, \sqrt{-m/t}\;,
\end{equation}
where the spectral parameter $\tilde{\mu}$ does not depend on time. We
extend the above equations to the full star diameter $-\infty < \zeta
< +\infty$ with $\chi_*(-\zeta) = -\chi_*(\zeta)$, as explained in
Appendix~\ref{sec:spherical-symmetry}.

\begin{figure}
  \includegraphics{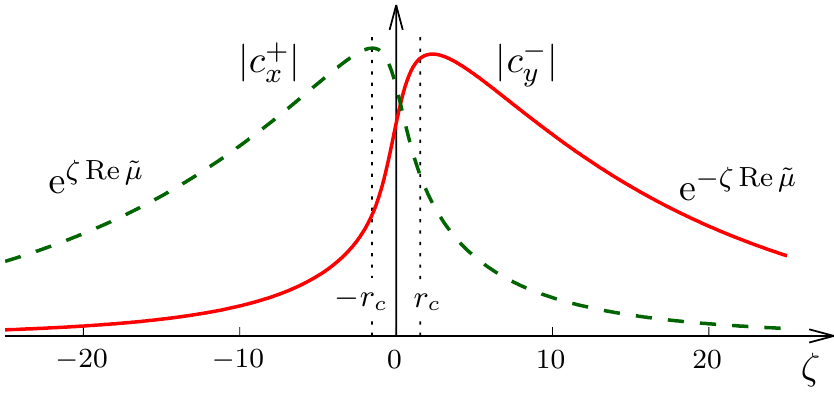}
  \caption{Resonance mode in the collapsing
    star; functions $c_x^+(\zeta)$ and $c_y^-(\zeta)$ are not
     symmetric to each other. We use  self-similar coordinate $\zeta$
     and $g_{a\gamma\gamma} = 0.37\, g_4/f_a$. The  respective
     eigenvalue is $\tilde{\mu} \approx
     0.065 + 0.025  \, i$.\label{fig:resonance-mode}}
\end{figure}

We numerically solve Eqs.~(\ref{eq:56}) with boundary
conditions~\eqref{bound}; the exemplary solution at $g'/g_4\approx 0.13$ is
shown in Fig.~\ref{fig:resonance-mode}. Notably, the nontrivial part of
this solution has width corresponding to $r_c(t)$ (vertical lines in
Fig.~\ref{fig:resonance-mode}). Beyond this part $|c_i^{\pm}|$ freely
decay as $\exp\{-|\zeta| \, \mathrm{Re}\, \tilde{\mu}\}$.  
Thus, the resonance mode shrinks on par with the collapsing star.

Numerical solutions of Eqs.~(\ref{eq:56}), exist only at 
\begin{equation}
  \label{eq:59}
  g_{a\gamma\gamma} \geq 0.25 \, \frac{g_4}{f_a}\;.
\end{equation}
This is a general condition to ignite parametric instability in
collapsing stars. It reproduces minimal coupling required for the
resonance in Fig.~\ref{fig:gmin} (horizontal dashed line).  Also, it
is twice weaker than the condition for critical stars before collapse,
cf.\ Eq.~\eqref{eq:13}. For QCD axions, the region~(\ref{eq:59}) is
above the dashed line in Fig.~\ref{fig:exclusion}.

\begin{figure}
  \unitlength=1mm
  \begin{picture}(85,51.5)
    \put(0,0){\includegraphics{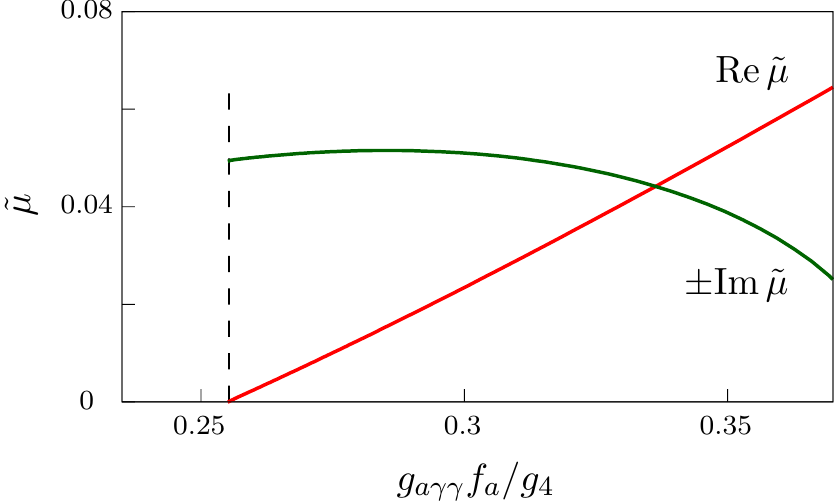}}
    \put(17,44){Eq.~\eqref{eq:59}}
  \end{picture}
  \caption{Rescaled growth exponents $\tilde{\mu}$ in the collapsing star.}
  \label{fig:mu-tilde}
\end{figure}

If the above inequality is met, the resonance progresses with two
complex time-dependent exponents $\mu$ and $\mu^*$ in
Eq.~(\ref{eq:55}), where $\pm\mathrm{Im}\, \mu$ are the Doppler
shifts. The respective eigenvalues $\tilde{\mu}$ are plotted in
Fig.~\ref{fig:mu-tilde}. Importantly, the time dependence of $\mu$
does not stop the resonance. Indeed, we already argued that the
respective mode behaves like a localized level in quantum
mechanics. Slow variations of external background do not change
occupation of this level if the adiabatic condition is satisfied, 
\begin{equation}
  \label{eq:31}
  \frac{\partial_t \mu}{\mu^2} \sim (-mt)^{-1/2} \ll 1\;.
\end{equation}
Thus, the electromagnetic field sits on two quasi-stationary resonance
levels,
\begin{equation}
  \notag
  C_{\alpha}^{\pm} = A\, c_{\alpha}^{\pm}(t,\, z)\, \mathrm{e}^{\int_{t_0}^t dt \,
    \mu } \pm A'\, \epsilon_{\alpha\beta}  \, [c_\beta^{\mp}(t,\, -z)]^*
  \, \mathrm{e}^{\int_{t_0}^t dt \, \mu^*},
\end{equation}
at least until the backreaction ruins the self-similar background. 

\begin{figure}
  \unitlength=1mm
  \begin{picture}(85,51.5)
    \put(0,0){\includegraphics{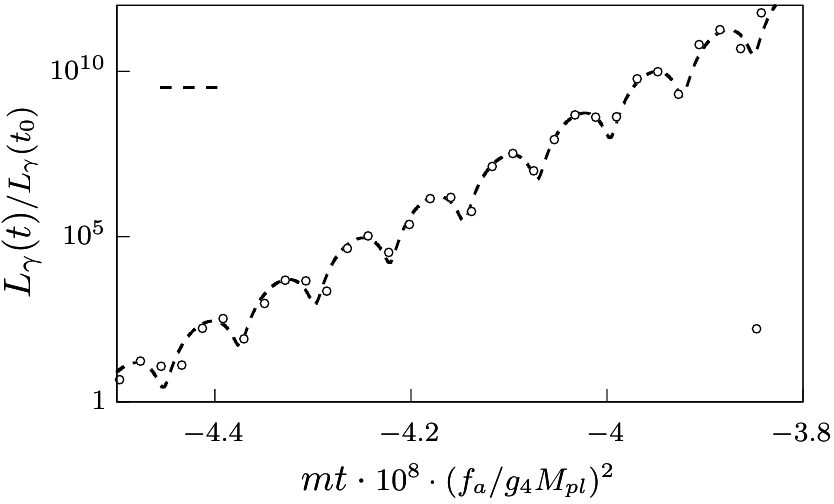}}
    \put(47,17){Eqs.~\eqref{sys} and \eqref{eq:69}}
    \put(25.5,41.5){Eqs.~(\ref{eq:55}) and \eqref{eq:69}}
  \end{picture}
  \caption{Luminosity~(\ref{eq:69}) of parametric emission from the
    collapsing star in Fig.~\ref{fig:collapse}a at $g_{a\gamma\gamma}
    = 0.33\, g_4/f_a$, $b=0.9$, and $\varphi_0 = 0$. Self-similar
    result~(\ref{eq:55}) (dashed line) is compared to the direct solution
    of Eqs.~\eqref{sys} (points). \label{fig:nonrelativistic_linear}}
\end{figure}

The axion star radio-luminosity follows from the above
representation. Interestingly, it oscillates in time due to
interference between the modes, 
\begin{equation}
  \label{eq:69}
  L_\gamma \propto \mathrm{e}^{2\mathrm{Re} \int_{t_0}^t \mu\, dt}
  \left [1 + b\, \cos\left(2 \mathrm{Im} \int_{t_0}^t \mu \, dt +
    \varphi_0 \right) \right],
\end{equation}
where $b$ and $\varphi_0$  depend on the initial amplitudes $A$,
$A'$, with $b=1$ representing equipartition.
In Fig.~\ref{fig:nonrelativistic_linear} we
illustrate\footnote{For simplicity we ignore time dependence of
    the resonance wave functions.} these
oscillations at $b=0.9$, $\varphi_0 = 0$. Dashed line in 
this figure represents self-similar formula with $\int \mu \,
dt = -2\tilde{\mu} (-mt)^{1/2}$. It coincides with the direct result
(points) obtained by solving Eqs.~\eqref{sys} for $\mu(t)$ numerically
in the background of a collapsing star and then using  
Eq.~(\ref{eq:69}). This supports our analytic solution in
Eq.~(\ref{eq:55}). 

\begin{figure}
  \unitlength=1mm
  \begin{picture}(85,65)
    \put(0,0){\includegraphics{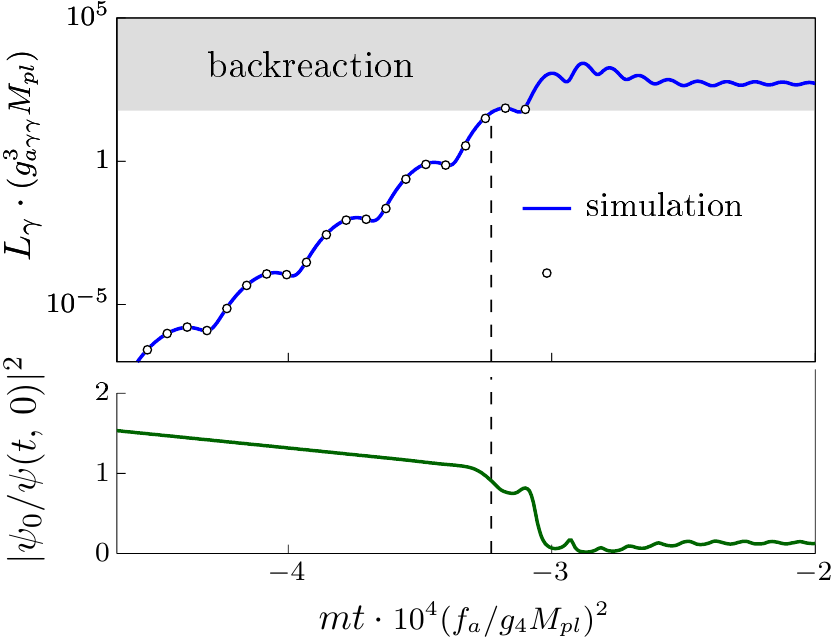}}
    \put(59,36.5){\small Eqs.~\eqref{sys}, \eqref{eq:69}}
  \end{picture}
  \caption{Luminosity $L_{\gamma}(t)$ of critical collapsing star during parametric
    resonance;     $g_{a\gamma\gamma} = 0.33\, g_4/f_a$. Full
    numerical simulation (solid line) is compared to the solution of
    Eqs.~\eqref{sys} (points). We use universal units from 
    Appendix~\ref{sec:comp-with-bose}; in particular, $\psi_0^2 =
    10^{3} \, (f_a / g_4^2 M_{pl})^2$. \label{fig:flux-collapsing}} 
\end{figure}

To test the above picture of parametric resonance during collapse, we
simulate the coupled system of relativistic 
equations~\eqref{equation}, \eqref{eq:5} for photons and axions, see
Fig.~\ref{fig:flux-collapsing}, movie~\cite{movie}, and 
Appendix~\ref{sec:full-relat-simul} for details. We find that at
first, the star  squeezes with no effect on the electromagnetic
field. But once the localized solution of Eqs.~\eqref{sys} appears,
growth and oscillations of the luminosity begin (solid line in
Fig.~\ref{fig:flux-collapsing}). The exact result is reproduced by 
Eq.~(\ref{eq:69}) (points), where $\mu(t)$ is obtained by solving the
boundary value problem~\eqref{sys} and $b$, $\phi_0$ are obtained from
the fit.

  It is worth reminding that Eq.~\eqref{eq:69} is applicable only for
  nonrelativistic stars deep in the self-similar regime. This is
  possible only at very large values of $mR_s$ which are hard to
  achieve in relativistic simulations. In particular, the value of
  $\mu$ in Eq.~\eqref{eq:55} is by a factor of two different from the
  simulation in Fig.~\ref{fig:flux-collapsing}.

We finish this Section with a
mystery. Figure~\ref{fig:flux-collapsing} demonstrates that once the
inequality~\eqref{eq:45} is broken (shaded region), the backreaction
ruins self-similar dynamics. Indeed, the axion field\footnote{In relativistic
  simulation $|\psi| \equiv |\partial_t a - im a|/(f_a m \sqrt{2})$, see
  Eq.~\eqref{ansatz}.} does not behave
anymore as $|\psi(t,\, 0)|^{-2}  \propto -t$, like Eq.~\eqref{eq:sscol}
suggests. Nevertheless, the luminosity continues to grow and saturates
only deep inside the 
backreaction region. We will investigate this nonlinear regime in the
forthcoming publication~\cite{Levkov}. 

For QCD axions, the saturated luminosity in
Fig.~\ref{fig:flux-collapsing}~is, 
\begin{equation}
L_\gamma = 1.5\cdot 10^{41} \left(\frac{m}{26\, \mu \mathrm{eV}}
\right)^{-3}  {\rm erg} \cdot {\rm s}^{-1} \;,
\label{eq:Lum_steady}
\end{equation}
while the corresponding flux strongly depends on direction, see
Fig.~\ref{fig:chaos}. Notably, this is close to the parameters of
Fast Radio Bursts, $L_{\mathrm{FRB}} = 10^{38} - 10^{40}\,
\mathrm{erg} \cdot \mathrm{s}^{-1}$. 

\section{Discussion}
\label{sec:discussion}
In this paper we have found that the finite-volume parametric
resonance is described by the quasi-stationary Schr\"odinger-like
system~\eqref{sys} with non-Hermitean  ``Hamiltonian.'' That is where
the fun has begun! Photon instability modes became localized states,
and their growth 
exponents ${\mathrm{Re} \, \mu > 0}$~---  
eigenvalues of the ``Hamiltonian.'' The condition for the resonance then
indicates whether the localized states exist. Using this technique, we
computed the resonance condition for the isolated Bose stars, collapsing
and moving stars, their groups, and diffuse axions. We  argued that
axions with relative velocities exceeding a certain value of order
$(mR)^{-1}$, are sharply out of resonance, where $R$ is the system
size.

With help of quantum-mechanical perturbation theory, we analytically
computed the instability modes and growth exponents in the physically 
motivated case of slow resonance, $\mu R \ll 1$. Interestingly, our
theory predicts a long-living quasi-stationary photon mode
with small negative decay exponent $\mathrm{Re}\, \mu<0$ after the 
resonance switches off, and we see this mode in simulations. 

We have found two unexpected applications of our method. First, it
describes stimulated emission of ambient radiation in axion stars. We
observed that these objects can realistically give larger contribution
to the radiobackground than the diffuse axions, producing a thin
spectral line at $\omega_{\gamma} \approx m/2$. Second, with
additional coarse-graining our approach reproduces well-known kinetic
equation for photons interacting with virialized axions. 

A warning is in order: our technique is applicable only in the
  case of nonrelativitic axions at high occupation numbers. These
  approximations may break down only under extreme conditions, say, in the
  strong gravitational field of a black hole or a neutron star, or at
  very late stages of Bose star collapse. That is why our method should
  work in vast majority of astrophysical settings with dark matter
  axions, and we expect that truly cool applications are still ahead. Besides,
astrophysics offers an impressive set of situations  
where the resonance condition can be satisfied, and the ones with the
largest $D_{\infty}$ are of primary interest. Using our method, one
can study parametric instability in superradiant axion clouds near
rotating black holes~\cite{Rosa:2017ury}, or in tidally elongated
axion stars falling onto the neutron stars~\cite{Tkachev:2014dpa}, or
in groups of gravitationally bound Bose
stars~\cite{Hertzberg:2018zte}. In all these cases an observable
radio-flash can appear, constraining the  axion models or even
explaining Fast Radio Bursts~\cite{Petroff:2019tty}. On the calmer side,
objects at the rim of parametric resonance can give large
contributions into the radiobackground possibly explaining
ARCADE~2~\cite{Fixsen:2009xn} and  EDGES~\cite{Bowman:2018yin} anomalies.

Technically, we completely disregarded potentially important
light-bending and divergence effects of the resonance rays,
cf.~\cite{Blas:2019qqp,McDonald:2019wou}, as well as phenomena of
astrophysical plasma.  These certainly deserve a separate study. 

We explicitly saw that gravitational and self-interaction energies of
  axions inside the star trivially shift the photon frequencies 
without affecting the resonance. We do not expect these effects to be
important in other situations as well. In particular, the distribution
function of virialized axions in the galaxy depends on their total
energy $E$, not kinetic or potential. The photon of frequency
$\omega_\gamma \approx E/2$ will stay in resonance with same part of
the ensemble in different parts of the 
galaxy~\cite{Tkachev:1987cd,Riotto:2000kh}. Thus, the main  
show-stoppers for the parametric instabilities are the Doppler shifts and
backreaction effects.


\acknowledgments

We are indebted to Elena Sokolova for encouraging interest. 
We thank all participants of the MIAPP-2020 program ``Axion
  Cosmology'' for discussions. Work on parametric resonance in Bose 
  stars was
  supported by the grant RSF 16-12-10494. The rest of this paper
  received support from the Foundation for the Advancement of
  Theoretical Physics and Mathematics ``BASIS'' and the Munich Institute   
  for Astro- and Particle Physics (MIAPP), funded by the Deutsche
  Forschungsgemeinschaft under Germany's Excellence Strategy~---
  EXC-2094–390783311. Numerical calculations were performed on the
  Computational cluster of the Theory Division of INR~RAS.


\appendix

\section{Spherically-symmetric case}
\label{sec:spherical-symmetry}

In the background of a spherical axion star with ${\psi = \psi(t,\, r)}$
it is natural to decompose the electromagnetic field $\boldsymbol{A} =
\{A_i\}$ into spherical harmonics, 
\begin{equation}
  \label{elm}
  \boldsymbol{A} = \sum_{lm'} \left( A_Y^{lm'} \boldsymbol{Y}_{lm'} +
  A_\Psi^{lm'} \boldsymbol{\Psi}_{lm'} + 
  A_\Phi^{lm'} \boldsymbol{\Phi}_{lm'}\right)
\end{equation}
where we use the gauge $A_0 = 0$, spherical vectors
${\boldsymbol{Y}_{lm'} = \boldsymbol{x} \, Y_{lm'} /r}$,
${\boldsymbol{\Psi}_{lm'} = r \boldsymbol{\nabla}\, Y_{lm'}}$,  
${\boldsymbol{\Phi}_{lm'} = [\boldsymbol{\nabla}\times {\bf x}]\,
  Y_{lm'}}$, and denote the standard spherical functions by
  $Y_{lm'}(\theta, \, \phi)$. Below we omit the superscripts $lm'$
  for brevity. 

The coefficients of decomposition $A_{Y,\Psi, \Phi}(t,\, r)$ depend 
only on time $t$ and radial coordinate $r$. Substituting Eq.~(\ref{elm}) 
into the Maxwell's equation~\eqref{equation}, one finds the Gauss law
\begin{equation}
  \label{f2}
  A_\Psi = \frac{\partial_r(r^2A_Y)}{r \, l(l+1)}
\end{equation}
and two dynamical equations
\begin{subequations}
  \label{eq:51}
  \begin{align}
    \label{eq:48}
    &r^2 \partial_t^2 A_Y = \partial^2_r(r^2 A_Y)
    - l(l+1)A_Y \\ & \qquad \qquad \qquad \qquad\qquad\notag 
    - g_{a\gamma\gamma} \,l (l+1)(\partial_t a) \,r A_\Phi\;,\\
    \label{eq:52}
    &r\partial_t^2A_\Phi = \partial^2_r(r A_\Phi)
    - l(l+1) \,  A_\Phi/r \\ & \qquad \qquad\qquad \qquad \notag
    - g_{a\gamma\gamma}\partial_t a
    \left[ A_Y  - \frac{\partial^2_r(r^2A_Y)}{l(l+1)} \right]\;,
  \end{align}
\end{subequations}
where we omitted terms with $\partial_r a$ because they are
suppressed by extra powers of $(mr)^{-1}$ and will not contribute into
equations for $C$'s. 

We finally introduce the eikonal ansatz,
\begin{align}
  \label{eq:47}
  &(m r)^2 \,A_Y  =   2 i l(l+1) \left\{  C_Y^+
  \mathrm{e}^{im(r+t)/2} \right. \\ \notag &\qquad\qquad\qquad \qquad
  \left.+ C_Y^-\mathrm{e}^{im(r-t)/2}\right\} +
   \mathrm{h.c.}\;,\\ 
   \notag
   & mr\, A_\Phi  =  C_\Phi^+\,  \mathrm{e}^{im(r+t)/2}
   + C_\Phi^-\, \mathrm{e}^{im(r-t)/2} + \mathrm{h.c.}
\end{align}
Using it in the above equations and omitting the
$(mr)^{-1}$ suppressed contributions, we find eikonal
equations~\eqref{syst} at $z=r>0$  for the unknowns $(C_Y^+,\,
C_\Phi^{-})$ in place of $(C_x^+,\, C_y^-)$, with the additional
term~\eqref{eq:46} representing derivatives with respect to the
spherical angles: ${\Delta_{\theta \phi} = -l(l+1)}$. The pair 
$(C_\Phi^+,\, -C_Y^-)$ satisfies the same equations.

There are two subtleties in the spherically-symmetric case. First, the
transverse polarizations $A_{\Phi}$ and ${A_{\Psi} \propto rA_Y}$ are
proportional to $r^{-1}$, see Eqs.~(\ref{f2}), (\ref{eq:47}). This
introduces $r^{-2}$ falloff of the electromagnetic flux $F_{\gamma,\, out}$ at
infinity and additional factors ${l(l+1)/(4\pi m^2 r^2)}$ in the
  backreaction terms of Eqs.~\eqref{eq:8},~\eqref{eq:9}.

Second,  proper boundary conditions should be imposed at  ${r=0}$. Solving
Eqs.~\eqref{eq:51} to the leading order at $r \ll R_s$,  we find
that $A_{\Phi}$ and $rA_Y$ are linear combinations of  the 
Bessel spherical functions $j_l(mr/2) \exp\{\pm imt/2\}$. The $mr
\gg 1$ asymptotics of the latter give boundary
conditions
\begin{equation}
  \notag
  C_Y^+ = (-1)^{l} \,  (C_Y^-)^*\;, \qquad  C_\Phi^+ = (-1)^{l+1} \,
  (C_\Phi^-)^*  
\end{equation}
at $r= 0$. 

Importantly, there is no need to solve Eqs.~\eqref{syst} on the
half-line $z=r>0$. Instead, we extend $C_\alpha^{\pm}$ to another half-line
using $C_x^+(z) = (-1)^{l} \, [C_Y^- (-z)]^*$ and ${C_y^-(z) =
(-1)^{l+1}\, [C_\Phi^+(-z)]^*}$ at $z = -r < 0$. After that $C_x^+$
and $C_y^-$ satisfy Eqs.~\eqref{syst} along the entire star diameter
$-\infty < z < +\infty$, and the boundary conditions at $r=0$.

\section{The spectrum of a symplectic operator}
\label{sec:prop-sympl-oper}

Consider the eigenvalue problem~(\ref{sys}) at real $\psi$. We denote
the $2\times 2$ operator in its right-hand side by
\begin{equation}
  \label{eq:11}
  \hat{\cal L} = \begin{pmatrix}
    \partial_z & i g' m\psi  \\[4px]
    -ig' m \psi & -\partial_z
   \end{pmatrix} .
\end{equation}
One can explicitly check that this
operator is symplectic, i.e.\ satisfies
\begin{equation}
  \label{eq:10}
  \hat{\Omega} \hat{\cal L} = \hat{\cal L}^\dag\hat{\Omega},
  \qquad \mbox{where} 
  \qquad \hat{\Omega} = \begin{pmatrix}0 & -i \\ i & 0\end{pmatrix}
\end{equation}
is a symplectic form.

Now, suppose $|\xi\rangle = (c_x^+, \; c_y^-)^T$ is the eigenmode of
$\hat{\cal L}$ satisfying the resonance boundary
conditions~\eqref{bound}. In this case the scalar product
\begin{equation}
  \label{eq:15}
\langle \xi | \hat{\Omega}| \xi\rangle = i \int d z \, \left(c_y^{-*} \,
c_x^+  - c_x^{+*} \, c_y^{-} \right)
\end{equation}
converges; below we fix normalization ${\langle \xi | \hat{\Omega} |\xi
  \rangle = 1}$. Then
\begin{equation}
  \label{eq:16}
\mu^* = \langle \xi | \hat{\Omega} \hat{\cal L}| \xi\rangle^\dag =
\langle \xi |\hat{\cal L}^\dag \hat{\Omega} | \xi\rangle^\dag = \mu\;,
\end{equation}
where in the last equality we used Eq.~(\ref{eq:10}). Thus, the
localized resonance modes of $\hat{\cal L}$ satisfying (\ref{bound})
have real $\mu$.

Note that the eigenmodes of $\hat{\cal L}$ with different eigenvalues
are orthogonal to each other in the sense of the scalar product
(\ref{eq:15}). Indeed, repeating the computation \eqref{eq:16} for
eigenvectors $|\xi_1\rangle$ and $|\xi_2\rangle$ with exponents
$\mu_1$ and $\mu_2$, we find
\begin{equation}
  \label{eq:17}
  \mu_2^*\,  \langle \xi_1 |\hat{\Omega} | \xi_2 \rangle^\dag =  \mu_1\, \langle \xi_1
  |\hat{\Omega} | \xi_2 \rangle^\dag\;, 
\end{equation}
which proves $\langle \xi_1 |\hat{\Omega} | \xi_2 \rangle=0$. Moreover,
one can argue that
the set of $\hat{\cal L}$ eigenmodes~---  the resonance ones and
the ones from the continuum spectrum~--- forms complete basis in
the space of bounded functions $c_x^{+}$ and~$c_y^{-}$.

With the above definitions we can develop a perturbation theory
for the spectrum of $\hat{\cal L}$. Indeed, suppose at $\psi=\psi_0(z)$
the operator $\hat{\cal L} = \hat{\cal L}_0$ has a normalized
eigenmode $|\xi_0\rangle$ with zero eigenvalue, $\hat{\cal L}_0
|\xi_0\rangle = 0$. At slightly different $\psi = \psi_0(z) + \delta
\psi(z)$ this operator receives variation $\delta \hat{\cal L} = -g'm\, \delta
\psi\, \hat{\Omega}$. In this case its resonance eigenmode $|\xi
\rangle = |\xi_0\rangle + |\delta \xi \rangle$ is close to
$|\xi_0\rangle$, and the respective eigenvalue $\mu$ is small. The
eigenvalue problem  $\hat{\cal L}|\xi\rangle = \mu |\xi\rangle$ takes
the form,
\begin{equation}
  \label{eq:18}
  \delta \hat{\cal L} |\xi_0\rangle + \hat{\cal L}_0 |\delta\xi
  \rangle = \mu | \xi_0\rangle\;,
\end{equation}
where we ignored quadratic terms in perturbations. The scalar
product with $|\xi_0\rangle$ gives,
\begin{equation}
  \label{eq:19}
  \mu = \frac{\langle \xi_0 | \hat{\Omega} \delta \hat{\cal L}|
    \xi_0\rangle} {\langle \xi_0 | \hat{\Omega} | \xi_0 \rangle} = -
  {g' m} \, \frac{\langle \xi_0 | \delta \psi |
    \xi_0\rangle} {\langle \xi_0 | \hat{\Omega} | \xi_0 \rangle}
\end{equation}
Using explicit solution (\ref{sol}) for $\xi_0$, we finally obtain
\begin{equation}
  \label{eq:43}
  \mu = {g' m} \; \frac{\int dz \, \left[\psi (t,\, {\bf x}) -
      \psi_0({\bf x}) \right]}{ \int dz
    \, \sin (2D_0)}\;.
\end{equation}
With~\eqref{Sinf} this expression reproduces Eq.~(\ref{eq:21})
from the main text.

\section{Scaling symmetry}
\label{sec:comp-with-bose}
We calculate parameters of Bose stars using scaling symmetry of the
Schr\"odinger-Poisson system~\eqref{eq:36}, \eqref{eq:40}. Consider first the model
without self-coupling, $g_4 = 0$. One finds that change of variables
\begin{subequations}
  \label{eq:37}
\begin{align}
  \label{eq:32}
&{\bf x} =
\lambda\, \tilde{\bf x}   \;, && t = m\lambda^2\,
\tilde{t}\;,\\
&\Phi = \frac{\tilde{\Phi}}{(m\lambda)^2}\;,
  && \psi =   \frac{M_{pl}\, \tilde{\psi}}{m^2\lambda ^2f_a}
\end{align}
\end{subequations}
with  arbitrary~$\lambda$ removes all constants from the
equations. This scaling allows us to map the model with arbitrary
  parameters to a reference one with~${\tilde{\psi}(0) = 1}$. We
  perform numerical calculations in tilded variables and then scale
  back to physical. Parameter $\lambda$ disappears in final answers,
  if one expresses it via the chosen Bose star characteristics, e.g. its
  mass,
\begin{equation}
  \label{eq:38}
  M_s = m^2 f_a^2\int d^3 {\bf x} \, |\psi_s|^2 = \tilde{M}_s\,
  \frac{M_{pl}^2}{\lambda m^2}\;,
\end{equation}
where $\tilde{M}_s \approx 3.9$ is computed numerically. Similarly,
the parameter~\eqref{Sinf} equals,
\begin{equation}
  \label{eq:39}
  D_{\infty} \approx 0.80 \, g_{a\gamma\gamma} \,\frac{M_{pl}}{\lambda m}\;.
\end{equation}
Using this approach, we obtain Eqs.~\eqref{eq:12}, \eqref{eq:44}. 

In models with $g_4 \neq 0$ the self-interaction can be ignored at $M \ll
M_{cr}$,  see Eq.~\eqref{eq:BSC},  and we are back to the above 
situation. Stars with $M \geq M_{cr}$ are unstable. In the
main text we mostly consider the critical star with ${M =  
M_{cr}}$. In this case one excludes all parameters
from the equations using Eqs.~\eqref{eq:37} with $\lambda = g_4
M_{pl}/mf_a$, computes the critical star numerically, and then
restores the physical parameters. The integral~\eqref{Sinf} in this case equals
\begin{equation}
  \label{eq:42}
  D_{\infty} \approx 3.04 \; \frac{g_{a\gamma \gamma}f_a}{g_4}
\end{equation}
implying~\eqref{eq:13}. These ``self-interaction'' units are
  exploited in
Figs.~\ref{fig:collapse}, \ref{fig:gmin},
\ref{fig:nonrelativistic_linear}, \ref{fig:flux-collapsing}.

Finally, if self-coupling is  negligible but backreaction
of photons on axions is relevant,  all constants can be eliminated from
Eqs.~\eqref{sys},~\eqref{eq:8}, (\ref{eq:40}) using Eqs.~(\ref{eq:37}),
$C_\alpha^{\pm} = \tilde{C}_\alpha^{\pm} (M_{pl}/g_{a\gamma\gamma})^{1/2}
(m\lambda)^{-2}$,  $\mu = \tilde{\mu}/\lambda$, and  ${\lambda =
g_{a\gamma \gamma} M_{pl}/m}$. We perform this rescaling to plot
universal quantities in
Figs.~\ref{fig:linear}, \ref{fig:mu1}, and~\ref{fig:flux-collapsing}.

\section{Initial conditions}
\label{sec:quantum-start}

In real astrophysical settings the axion stars are embedded into the
background of classical radiowaves which can give a good initial kick
to the parametric instability,
cf.\ Sec.~\ref{sec:light-amplification}. But this mechanism 
essentially depends on the environment, so outside of
  Sec.~\ref{sec:light-amplification} we assume quantum start,
  i.e.\ the resonance set off by the spontaneous decays of axions inside
  the isolated star.

Detailed study of quantum evolution is beyond the scope of this
  paper, so we use a shortcut.
Namely, the flux $F_{\gamma} \sim
  |\boldsymbol{E}|^2 \sim |\boldsymbol{H}|^2$ of spontaneous photons
  can be estimated from energy conservation,
\begin{equation}
  \label{flux}
  \partial_t M_s  = - \Gamma_{a\gamma\gamma} M_{s} = - 4\pi
  r^2 F_{\gamma}\;,
\end{equation}
where we assumed spherical Bose star and introduced the axion decay width
$\Gamma_{a\gamma\gamma} = g_{a\gamma\gamma}^2m^3/64 \pi$. This gives
typical amplitudes  
\begin{equation}
  \label{back}
  |\boldsymbol{E}| \sim |\boldsymbol{H}| \sim  \frac{1}{R_s}
  \left(
  \frac{M_{s}\Gamma_{a\gamma\gamma}}{4 \pi}\right)^{1/2}
\end{equation}
of spontaneous emission.

It is worth reminding  that the exponential growth of the resonance mode
washes out all details of initial quantum evolution, with just one
logarithmically sensitive parameter surviving: the time of
growth. That is why the above order-of-magnitude description is
adequate.  

In numerical simulation of Appendix~\ref{sec:full-relat-simul} we
mimic the quantum bath of spontaneous 
photons using a stochastic ensemble of random classical waves with
amplitudes~\eqref{back}. This is required only in dynamical
situations such as the axion star collapse in Sec.~\ref{sec:collapse}.   

\section{Full relativistic simulation}
\label{sec:full-relat-simul}

We test the theory by numerically evolving the equations~(\ref{equation}) and 
(\ref{eq:5}) for the electromagnetic and axion fields. In computations
we consider
only spherically symmetric  axion backgrounds, $a = a(t,\, r)$. This
is justified at the linear stages of parametric resonance 
and should be valid at least\footnote{The backreaction stage in the
  central part of  
  Fig.~\ref{fig:linear} is short, and related asphericities should 
  be small. Self-similar evolution in Fig.~\ref{fig:flux-collapsing}
  tracks spherically-symmetric attractor which suppresses axion modes
  with nonzero $l$.}
qualitatively during backreaction.  
To make Eq.~\eqref{eq:5} self-consistent, we average its 
right-hand  side over spherical angles: $F_{\mu \nu}  
\tilde{F}_{\mu\nu} \to \int d\Omega \, F_{\mu \nu} \tilde{F}_{\mu\nu}
/4\pi$. We decompose electric and magnetic fields $E_i = F_{0i}$ and
$H_i = - \epsilon_{ijk} F_{jk}/2$ in spherical harmonics
$\boldsymbol{Y}_{lm'}$, $\boldsymbol{\Psi}_{lm'}$, and
$\boldsymbol{\Phi}_{lm'}$ introduced in
Appendix~\ref{sec:spherical-symmetry}. With the cutoff $l \leq
l_{max}$, we find  $6l_{max}(l_{max}+2) +1$
equations\footnote{Note that $l =0$ components of $\boldsymbol{E}$ and
  $\boldsymbol{H}$ are absent.} for the same number of unknowns
$E_{lm'}^{Y,\, \Psi,\, \Phi}(t,\, r)$,  $H_{lm'}^{Y,\, \Psi,\,
  \Phi}(t,\, r)$, and $a(t,\, r)$.

As usual, the longitudinal number $m'$ does not explicitly appear in
equations for the spherical components of $\boldsymbol{E}$ and
$\boldsymbol{H}$. We therefore leave only one component at every $l$
multiplying its contribution in the right-hand side of
Eq.~\eqref{eq:5} by ${(2l+1)}$. Now, the number  of equations is
$6l_{max}-5$.

In practice our numerical results are insensitive to $l_{max}$: the
photon modes evolve independently at the linear stage, 
while backreaction simply equidistributes energy over them\footnote{The
  time when the backreaction appears is logarithmically sensitive to
  $l_{\max}$, however, cf.~Eq.~\eqref{eq:45}.}. We therefore perform
simulations in Figs.~\ref{fig:linear},~\ref{fig:mu1},
\ref{fig:flux-collapsing} with $l_{max}=1$ and use
$l_{max}=210$ with step $\Delta l=4$ to find the angular structure of
  the resonance in Sec.~\ref{sec:prof-reson-mode}. We restore 
three-dimensional electromagnetic fields during linear evolution 
multiplying the spherical components with their harmonics, e.g.
$$
\boldsymbol{E} = \sum_{lm'} E_l^\Psi(t,\, r) \, e_{lm'} \,
\boldsymbol{\Psi}_{lm'}(\theta,\, \phi) + \dots\;.
$$
where the dots hide other polarizations and independent random numbers
$e_{lm'}$ mimic quantum distribution of the initial resonance
amplitudes over the longitudinal number $m'$, see
Appendix~\ref{sec:quantum-start}. 

To hold the axions together during resonance, we add interaction with
the gravitational potential by changing ${\cal V}' \to (1 + 2\Phi){\cal
  V}'$ in Eq.~\eqref{eq:5}. This approximation is trustworthy if the
gravitational field is mostly sourced  by the nonrelativistic axions.  

Since our simulations check nonrelativistic theory,
we perform them only for small-velocity axions. In physical units,
parameters of these simulations correspond to $m = 26\,
\mu\mbox{eV}$, $g_4 = 0.59$ or~$0$, with other
parameters ranging in wide intervals
${f_a^2 = (10^{-11}\div 10^{-8})\, M_{pl}^2}$,~~${g_{a\gamma\gamma} =
  (0.15\div 0.4)\, f_a^{-1}}$, and $M_s = (10^{-11} \div 10^{-8}) \, M_{\odot}$. This indeed 
corresponds to small nonrelativistic parameter $(m
R_s)^{-1} = 10^{-3} \div10^{-6}$. Note that in universal units of
Figs.~\ref{fig:linear}, \ref{fig:mu1}, \ref{fig:angular_distribution},
\ref{fig:chaos}, 
\ref{fig:flux-collapsing} the results of our simulations look the same at essentially
different parameters. 

We store $a(t,\, r)$, $\Phi(t,\, r)$, and the components of
$\boldsymbol{E}$, $\boldsymbol{H}$ on a uniform radial lattice with
$\Delta r = 1.3/m$, 
using Fourier transform to compute their $r$-derivatives in
Eqs.~(\ref{equation}), (\ref{eq:5}), \eqref{eq:40}. Time evolution is
then performed with the fourth-order Runge-Kutta integrator with ${\Delta t
= 0.025/m}$. Equation~\eqref{eq:40} is solved at each step. In our
calculations the total energy is conserved at the level of~$10^{-8}$.

In the beginning of simulation we evolve the axion field alone, checking
Eqs.~\eqref{sys} for the resonance mode ($\mathrm{Re}\,\mu > 
0$) to appear. Once it is there\footnote{If not, the photon waves
  trivially leave the axion star.}, we randomly populate the Fourier 
modes of the electromagnetic field in the narrow frequency band 
$\omega_\gamma \approx m/2$, with typical amplitude~(\ref{back}) in
the $r$-space. This sets off the resonance making $\boldsymbol{E}$ and
$\boldsymbol{H}$ grow. 

We absorb the electromagnetic emission by introducing the ``Hubble''
friction at the lattice boundary $r>r_1$. The outgoing luminosity 
$L_{\gamma} = r^2\int d\Omega \,\boldsymbol{n}_r [\boldsymbol{E}\times
  \boldsymbol{H}]$ is measured at~${r = r_1}$. 

In Figs.~\ref{fig:collapse} and \ref{fig:nonrelativistic_linear} we
use the code of Ref.~\cite{Levkov:2016rkk} to evolve the
Schr\"odinger-Poisson equations~\eqref{eq:36}, \eqref{eq:40} for axions. 
Backreaction of photons on axions is not taken into account in these
calculations.



\begin{thebibliography}{99} 
\bibitem{Kim:2008hd} 
  J.~E.~Kim and G.~Carosi,
  \href{https://doi.org/10.1103/RevModPhys.82.557}{Rev.\ Mod.\ Phys.\  \textbf{82},
    557 (2010)}
       [\href{https://doi.org/10.1103/RevModPhys.91.049902}{{\it
             erratum-ibid} \textbf{91}, 049902 (2019)}]
  [\href{https://arxiv.org/pdf/0807.3125}{0807.3125}].

\bibitem{pdg}
  A.~Ringwald, L.~J.~Rosenberg, and G.~Rybka, ``Axions and other similar
  particles'', In
  \href{http://pdg.lbl.gov/2019/reviews/rpp2019-rev-axions.pdf}{Review
    of Particle Physics, 2020} (Particle Data Group).

\bibitem{Sikivie:2006ni}
  P.~Sikivie,
  \href{https://doi.org/10.1007/978-3-540-73518-2_2}{Lect.\ Notes
  Phys.\  \textbf{741}, 19-50 (2008)}
  [\href{https://arxiv.org/pdf/astro-ph/0610440}{astro-ph/0610440}].

\bibitem{Arias:2012az}
P.~Arias, D.~Cadamuro, M.~Goodsell, J.~Jaeckel, J.~Redondo and
A.~Ringwald,
\href{https://doi.org/10.1088/1475-7516/2012/06/013}{JCAP \textbf{06},
  013 (2012)}
[\href{https://arxiv.org/pdf/1201.5902}{arXiv:1201.5902}].
  
\bibitem{Peccei:1977hh} 
  R.~D.~Peccei and H.~R.~Quinn,
  \href{https://doi.org/10.1103/PhysRevLett.38.1440}{Phys.\ Rev.\ Lett.\  {\bf
      38}, 1440 (1977).}

\bibitem{Arvanitaki:2009fg} 
  A.~Arvanitaki, S.~Dimopoulos, S.~Dubovsky, N.~Kaloper and
  J.~March-Russell,
  \href{https://doi.org/10.1103/PhysRevD.81.123530}{Phys.\ Rev.\ D
    {\bf 81}, 123530 (2010)} [\href{https://arxiv.org/pdf/0905.4720}{0905.4720}]

\bibitem{diCortona:2015ldu} 
  G.~Grilli di Cortona, E.~Hardy, J.~Pardo Vega and G.~Villadoro,
  \href{https://doi.org/10.1007/JHEP01(2016)034}{JHEP {\bf 1601}, 034 (2016)}
  [\href{https://arxiv.org/pdf/1511.02867}{1511.02867}].

\bibitem{Irastorza:2018dyq} 
  I.~G.~Irastorza and J.~Redondo,
  \href{https://doi.org/10.1016/j.ppnp.2018.05.003}{Prog.\ Part.\ Nucl.\ Phys.\  {\bf
      102}, 89 (2018)}
  [\href{https://arxiv.org/pdf/1801.08127}{1801.08127}].

\bibitem{Armengaud:2019uso} 
  E.~Armengaud {\it et al.} (IAXO Collaboration),
  \href{https://doi.org/10.1088/1475-7516/2019/06/047}{JCAP {\bf 1906}, 047 (2019)}
       [\href{https://arxiv.org/pdf/1904.09155}{1904.09155}].

\bibitem{Arvanitaki:2014wva} 
  A.~Arvanitaki, M.~Baryakhtar and X.~Huang,
  \href{https://doi.org/10.1103/PhysRevD.91.084011}{Phys.\ Rev.\ D
    {\bf 91}, no. 8, 084011 (2015)}
  [\href{https://arxiv.org/pdf/1411.2263}{1411.2263}].
       
\bibitem{Arza:2019nta}
A.~Arza and P.~Sikivie,
\href{https://doi.org/10.1140/epjc/s10052-019-6759-7}{Phys.\ Rev.\ Lett.\  \textbf{123},
  131804 (2019)} [\href{https://arxiv.org/pdf/1902.00114}{1902.00114}].
  
\bibitem{Foster:2020pgt} 
  J.~W.~Foster {\it et al.},
  \href{http://arxiv.org/pdf/2004.00011.pdf}{2004.00011}.

\bibitem{Tkachev:1987cd} 
  I.~I.~Tkachev,
  \href{https://doi.org/10.1016/0370-2693(87)91318-9}{Phys.\ Lett.\ B
    {\bf 191}, 41 (1987).}

\bibitem{Preskill:1982cy} 
  J.~Preskill, M.~B.~Wise and F.~Wilczek,
  \href{https://doi.org/10.1016/0370-2693(83)90637-8}{Phys.\ Lett.\  {\bf
      120B}, 127 (1983)}.
  
\bibitem{Abbott:1982af} 
  L.~F.~Abbott and P.~Sikivie,
  \href{https://doi.org/10.1016/0370-2693(83)90638-X}{Phys.\ Lett.\  {\bf
      120B}, 133 (1983)}.
  
\bibitem{Alonso-Alvarez:2019ssa} 
  G.~Alonso-Álvarez, R.~S.~Gupta, J.~Jaeckel and M.~Spannowsky,
  \href{https://arxiv.org/pdf/1911.07885}{1911.07885}.

\bibitem{Riotto:2000kh} 
  A.~Riotto and I.~Tkachev,
  \href{https://doi.org/10.1016/S0370-2693(00)00660-2}{Phys.\ Lett.\ B
    {\bf 484}, 177 (2000)}
   [\href{https://arxiv.org/pdf/astro-ph/0003388}{astro-ph/0003388}].
  
\bibitem{Kephart:1994uy} 
  T.~W.~Kephart and T.~J.~Weiler,
  \href{https://doi.org/10.1103/PhysRevD.52.3226}{Phys.\ Rev.\ D {\bf
      52}, 3226 (1995)}.
 
\bibitem{Tkachev:2014dpa} 
  I.~I.~Tkachev,
  \href{https://doi.org/10.1134/S0021364015010154}{JETP Lett.\  {\bf
      101}, 1 (2015)}
  [Pisma Zh.\ Eksp.\ Teor.\ Fiz.\  {\bf 101}, 3 (2015)]
  [\href{https://arxiv.org/pdf/1411.3900}{1411.3900}].

\bibitem{Hertzberg:2018zte} 
  M.~P.~Hertzberg and E.~D.~Schiappacasse,
  \href{https://doi.org/10.1088/1475-7516/2018/11/004}{JCAP {\bf
      1811}, 004 (2018)}
  [\href{https://arxiv.org/pdf/1805.00430}{1805.00430}].

\bibitem{Arza:2018dcy} 
  A.~Arza,
  \href{https://doi.org/10.1140/epjc/s10052-019-6759-7}{Eur.\ Phys.\ J.\ C
    {\bf 79}, 250 (2019)}
  [\href{https://arxiv.org/pdf/1810.03722}{1810.03722}].
  
\bibitem{Sigl:2019pmj} 
  G.~Sigl and P.~Trivedi,
  \href{https://arxiv.org/pdf/1907.04849}{1907.04849}.
 
 \bibitem{Carenza:2019vzg} 
  P.~Carenza, A.~Mirizzi and G.~Sigl,
  \href{https://arxiv.org/pdf/1911.07838}{1911.07838}.
  
\bibitem{Chen:2020ufn} 
  L.~Chen and T.~W.~Kephart,
  \href{https://arxiv.org/pdf/2002.07885}{2002.07885}.

\bibitem{Wang:2020zur} 
  Z.~Wang, L.~Shao and L.-X.~Li,
  \href{https://arxiv.org/pdf/2002.09144}{2002.09144}.

\bibitem{Arza:2020eik}
A.~Arza, T.~Schwetz and E.~Todarello,
[\href{https://arxiv.org/pdf/2004.01669}{2004.01669}].
  
\bibitem{Patras2018} 
D.~G.~Levkov, A.~G.~Panin and I.~I.~Tkachev, ``Laser effect for cosmic
axions,''
\href{https://indico.desy.de/indico/event/20012/session/23/contribution/65}{14th
  Patras Workshop on Axions, WIMPs and WISPs, DESY, 
Hamburg}, June 18 - 22, 2018
     [\href{https://indico.desy.de/indico/event/20012/session/23/contribution/65/material/poster/0.pdf}{poster},
       \href{https://indico.desy.de/indico/event/20012/session/23/contribution/65/material/slides/0.pdf}{presentation}]. 

\bibitem{Patras2019} 
D.~G.~Levkov, A.~G.~Panin and I.~I.~Tkachev, ``Collapsing Bose
  stars as source of repeating fast radio bursts,''
  \href{https://indico.desy.de/indico/event/22598/session/22/contribution/36}{15th
    Patras Workshop on Axions, WIMPs and WISPs}, Freiburg, June 3-7 ,
  2019
  [\href{https://indico.desy.de/indico/event/22598/session/22/contribution/36/material/slides/0.pdf}{presentation}].

\bibitem{Dine:1982ah}
M.~Dine and W.~Fischler,
\href{https://doi.org/10.1016/0370-2693(83)90639-1}{Phys. Lett. B
  \textbf{120}, 137 (1983)}.
  
\bibitem{Niemeyer:2019aqm} 
  J.~C.~Niemeyer,
  \href{http://arxiv.org/pdf/1912.07064.pdf}{arXiv:1912.07064}.
  
\bibitem{Kolb:1993hw} 
  E.~W.~Kolb and I.~I.~Tkachev,
  \href{https://doi.org/10.1103/PhysRevD.49.5040}{Phys.\ Rev.\ D {\bf
      49}, 5040 (1994)}
  [\href{https://arxiv.org/pdf/astro-ph/9311037}{astro-ph/9311037}].

\bibitem{Klaer:2017ond} 
  V.~B.~Klaer and G.~D.~Moore,
  \href{https://doi.org/10.1088/1475-7516/2017/11/049}{JCAP {\bf
      1711}, 049 (2017)}
  [\href{https://arxiv.org/pdf/1708.07521}{1708.07521}].

\bibitem{Gorghetto:2018myk}
M.~Gorghetto, E.~Hardy and G.~Villadoro,
\href{https://doi.org/10.1007/JHEP07(2018)151}{JHEP \textbf{07}, 151 (2018)}
[\href{https://arxiv.org/pdf/1806.04677}{1806.04677}].
  
\bibitem{Vaquero:2018tib} 
  A.~Vaquero, J.~Redondo and J.~Stadler,
  \href{https://doi.org/10.1088/1475-7516/2019/04/012}{JCAP {\bf
      1904}, 012 (2019)}
  [\href{https://arxiv.org/pdf/1809.09241}{1809.09241}].

  \bibitem{Buschmann:2019icd} 
  M.~Buschmann, J.~W.~Foster and B.~R.~Safdi,
  \href{https://arxiv.org/pdf/1906.00967}{arXiv:1906.00967}.  

\bibitem{Hogan:1988mp} 
  C.~J.~Hogan and M.~J.~Rees,
  \href{https://doi.org/10.1016/0370-2693(88)91655-3}{Phys.\ Lett.\ B
    {\bf 205}, 228 (1988)}.
  
\bibitem{Kolb:1993zz} 
  E.~W.~Kolb and I.~I.~Tkachev,
  \href{https://doi.org/10.1103/PhysRevLett.71.3051}{Phys.\ Rev.\ Lett.\  {\bf
      71}, 3051 (1993)}
  [\href{https://arxiv.org/pdf/hep-ph/9303313}{hep-ph/9303313}].

\bibitem{Kolb:1994fi} 
  E.~W.~Kolb and I.~I.~Tkachev,
  \href{https://doi.org/10.1103/PhysRevD.50.769}{Phys.\ Rev.\ D {\bf
      50}, 769 (1994)}
  [\href{https://arxiv.org/pdf/astro-ph/9403011}{astro-ph/9403011}].
  
 \bibitem{Eggemeier:2019khm} 
  B.~Eggemeier, J.~Redondo, K.~Dolag, J.~C.~Niemeyer and A.~Vaquero,
  \href{https://arxiv.org/pdf/1911.09417}{1911.09417}.

\bibitem{Ruffini:1969qy} 
  R.~Ruffini and S.~Bonazzola,
  \href{https://doi.org/10.1103/PhysRev.187.1767}{Phys.\ Rev.\  {\bf
      187}, 1767 (1969).}

  \bibitem{Tkachev:1986tr} 
  I.~I.~Tkachev,
  \href{https://ui.adsabs.harvard.edu/abs/1986SvAL...12..305T/abstract}{Sov.\ Astron.\ Lett.\  {\bf 12}, 305 (1986)
  [Pisma Astron.\ Zh.\  {\bf 12}, 726 (1986)].}

 \bibitem{Levkov:2018kau} 
  D.~G.~Levkov, A.~G.~Panin and I.~I.~Tkachev,
  \href{https://doi.org/10.1103/PhysRevLett.121.151301}{Phys.\ Rev.\ Lett.\  {\bf
        121}, no. 15, 151301 (2018)}
  [\href{https://arxiv.org/pdf/1804.05857}{arXiv:1804.05857}].

\bibitem{Eggemeier:2019jsu} 
  B.~Eggemeier and J.~C.~Niemeyer,
  \href{https://doi.org/10.1103/PhysRevD.100.063528}{Phys.\ Rev.\ D
    {\bf 100}, 063528 (2019)}
  [\href{https://arxiv.org/pdf/1906.01348}{1906.01348}].

\bibitem{Arvanitaki:2010sy}
  A.~Arvanitaki and S.~Dubovsky,
  \href{https://doi.org/10.1103/PhysRevD.83.044026}{Phys.\ Rev.\ D
    \textbf{83}, 044026 (2011)}
       [\href{https://arxiv.org/pdf/1004.3558}{1004.3558}].

\bibitem{Stott:2018opm}
  M.~J.~Stott and D.~J.~Marsh,
  \href{https://doi.org/10.1103/PhysRevD.98.083006}{Phys.\ Rev.\ D
    \textbf{98}, 083006 (2018)}
  [\href{https://arxiv.org/pdf/1805.02016}{1805.02016}].

\bibitem{Rosa:2017ury}
J.~G.~Rosa and T.~W.~Kephart,
\href{https://doi.org/10.1103/PhysRevLett.120.231102}{Phys.\ Rev.\ Lett.\  \textbf{120},
  231102 (2018)}
[\href{https://arxiv.org/pdf/1709.06581}{1709.06581}].

\bibitem{Schive:2014dra} 
  H.~Y.~Schive, T.~Chiueh and T.~Broadhurst,
  \href{https://doi.org/10.1038/nphys2996}{Nature Phys.\  {\bf 10},
    496 (2014)}
  [\href{https://arxiv.org/pdf/1406.6586}{1406.6586}].
  
\bibitem{Schive:2014hza} 
  H.~Y.~Schive, M.~H.~Liao, T.~P.~Woo, S.~K.~Wong, T.~Chiueh,
  T.~Broadhurst and W.-Y.~P.~Hwang,
  \href{https://doi.org/10.1103/PhysRevLett.113.261302}{Phys.\ Rev.\ Lett.\  {\bf
      113}, 261302 (2014)}
  [\href{https://arxiv.org/pdf/1407.7762}{1407.7762}].

  \bibitem{Amin:2019ums}
  M.~A.~Amin and P.~Mocz,
  \href{https://doi.org/10.1103/PhysRevD.100.063507}{Phys. Rev. D
    \textbf{100} (2019) 063507} 
  [\href{https://arxiv.org/pdf/1902.07261}{1902.07261}].
  
\bibitem{Zakharov12}
  V.~E.~Zakharov and E.~A.~Kuznetsov,
  \href{http://doi.org/10.3367/UFNe.0182.201206a.0569}{Phys.\ Usp.\ {\bf
      55}, 535 (2012)}.
  
\bibitem{Chavanis:2011zi} 
  P.~H.~Chavanis,
  \href{https://doi.org/10.1103/PhysRevD.84.043531}{Phys.\ Rev.\ D
    {\bf 84}, 043531 (2011)}
  [\href{http://arxiv.org/pdf/1103.2050.pdf}{1103.2050}].

\bibitem{Levkov:2016rkk} 
  D.~G.~Levkov, A.~G.~Panin and I.~I.~Tkachev,
  \href{https://doi.org/10.1103/PhysRevLett.118.011301}{Phys.\ Rev.\ Lett.\  {\bf
      118}, 011301 (2017)}
  [\href{https://arxiv.org/pdf/1609.03611}{1609.03611}].

\bibitem{Petroff:2019tty} 
  E.~Petroff, J.~W.~T.~Hessels and D.~R.~Lorimer,
  \href{https://doi.org/10.1007/s00159-019-0116-6}{Astron.\ Astrophys.\ Rev.\  {\bf
      27}, 4 (2019)}
  [\href{https://arxiv.org/pdf/1904.07947}{1904.07947}].

 \bibitem{Blas:2019qqp} 
  D.~Blas, A.~Caputo, M.~M.~Ivanov and L.~Sberna,
  \href{https://doi.org/10.1016/j.dark.2019.100428}{Phys.\ Dark Univ.\  {\bf 27}, 100428 (2020)}
  [\href{https://arxiv.org/pdf/1910.06128}{1910.06128}].

\bibitem{McDonald:2019wou} 
  J.~I.~McDonald and L.~B.~Ventura,
  \href{https://arxiv.org/pdf/1911.10221}{1911.10221}.

\bibitem{Veltmaat:2016rxo} 
  J.~Veltmaat and J.~C.~Niemeyer,
  \href{https://doi.org/10.1103/PhysRevD.94.123523}{Phys.\ Rev.\ D
    {\bf 94}, 123523 (2016)}
       [\href{https://arxiv.org/pdf/1608.00802}{1608.00802}]. 

\bibitem{Veltmaat:2018dfz} 
  J.~Veltmaat, J.~C.~Niemeyer and B.~Schwabe,
  \href{https://doi.org/10.1103/PhysRevD.98.043509}{Phys.\ Rev.\ D
    {\bf 98}, 043509 (2018)}
  [\href{https://arxiv.org/pdf/1804.09647}{1804.09647}].
       
\bibitem{Veltmaat:2019hou} 
  J.~Veltmaat, B.~Schwabe and J.~C.~Niemeyer,
  \href{https://arxiv.org/pdf/1911.09614}{1911.09614}.

\bibitem{Schwabe:2016rze} 
  B.~Schwabe, J.~C.~Niemeyer and J.~F.~Engels,
  \href{https://doi.org/10.1103/PhysRevD.94.043513}{Phys.\ Rev.\ D
    {\bf 94}, 043513 (2016)}
       [\href{https://arxiv.org/pdf/1606.05151}{1606.05151}].
  
\bibitem{Schive:2019rrw} 
  H.~Y.~Schive, T.~Chiueh and T.~Broadhurst,
  \href{https://arxiv.org/pdf/1912.09483}{1912.09483}.

\bibitem{Hook:2018dlk} 
  A.~Hook,
  \href{https://s3.cern.ch/inspire-prod-files-1/134a370fc91047f492644df2128a1008}{PoS
    TASI {\bf 2018}, 004 (2019)}
  [\href{https://arxiv.org/pdf/1812.02669}{1812.02669}].

\bibitem{DiLuzio:2020wdo} 
  L.~Di Luzio, M.~Giannotti, E.~Nardi and L.~Visinelli,
  \href{https://arxiv.org/pdf/2003.01100}{2003.01100}.

\bibitem{Agrawal:2017cmd} 
  P.~Agrawal, J.~Fan, M.~Reece and L.~T.~Wang,
  \href{https://doi.org/10.1007/JHEP02(2018)006}{JHEP {\bf 1802}, 006
    (2018)}
  [\href{https://arxiv.org/pdf/1709.06085}{1709.06085}].

\bibitem{LL3}
  L.D.~Landau, E.M.~Lifshitz, {\it Quantum Mechanics: Non-Relativistic
    Theory. Course on theoretical physics, Vol. 3}. Pergamon Press, 1958.

\bibitem{Caputo:2018vmy} 
  A.~Caputo, M.~Regis, M.~Taoso and S.~J.~Witte,
  \href{https://doi.org/10.1088/1475-7516/2019/03/027}{JCAP {\bf
      1903}, 027 (2019)}
  [\href{https://arxiv.org/pdf/1811.08436}{1811.08436}].

\bibitem{movie}
  \url{https://www.youtube.com/playlist?list=PLMxQF3HFStX1sn_tCRAXWMmF5cCeGoNG_}
  
\bibitem{Levkov} 
  D.~G.~Levkov, A.~G.~Panin, I.~I.~Tkachev,
  {\it to appear}.

\bibitem{Fixsen:2009xn}
  D.~Fixsen {\it et al}, 
  \href{https://doi.org/10.1088/0004-637X/734/1/5}{Astrophys.\ J.\  \textbf{734},
    5 (2011)}
       [\href{https://arxiv.org/pdf/0901.0555}{0901.0555}].

\bibitem{Bowman:2018yin} 
  J.~D.~Bowman, A.~E.~E.~Rogers, R.~A.~Monsalve, T.~J.~Mozdzen and
  N.~Mahesh,
  \href{https://doi.org/10.1038/nature25792}{Nature {\bf 555}, 67 (2018)}
  [\href{https://arxiv.org/pdf/1810.05912}{1810.05912}].
       
\end{thebibliography}
\end{document}